# Quantitative acoustic monitoring of ensembles of weakly nonlinear microbubble oscillations in optically inaccessible environments

Hohyun Lee,[1] Reza Pakdaman Zangabad,[1] Chulyong Kim,[1] Victor Menezes[1], Juyoung Park[3,4], F. Levent Degertekin,[1] and Costas Arvanitis[1,2*]

**Affiliations**
[1] Georgia Institute of Technology, G.W. Woodruff School of Mechanical Engineering, Atlanta, USA
[2] Georgia Institute of Technology and Emory University, Wallace H. Coulter Department of Biomedical Engineering, Atlanta, USA
[3] Department of Biomedical Device, Gachon University College of IT Convergence, 1342 Seongnam-daero, Sujeong-gu, Seongnam 13120, Republic of Korea
[4] Neumous Inc., Osongsaengmyeong 1-ro, Osong-eup, Heungdeok-gu, Cheongju 28160, Republic of Korea

*Corresponding author
Costas Arvanitis (costas.arvanitis@gatech.edu)

## Abstract

A growing class of ultrasound-mediated diagnostic and therapeutic technologies, including sonoporation and blood–brain barrier modulation, relies on microbubble contrast agents, where precise control of microbubble dynamics governs biological responses, efficiency, and safety. However, quantitative monitoring of microbubble oscillations in the stable, weakly nonlinear regime remains challenging, particularly in optically opaque and deep-tissue environments. Here, we introduce a linear acoustic wave propagation and superposition (LAWPS) framework that reconstructs microbubble radius–time dynamics directly from passively recorded acoustic emissions. By coupling Fourier-series representations of weakly nonlinear oscillations with linear monopole radiation theory, LAWPS extends classical monopole models to establish a reversible relationship between multi-frequency acoustic emissions and underlying radial bubble dynamics. Extending this framework to monodisperse microbubble ensembles, we derive optimal excitation and receive configurations and population-level correction factors that enable quantitative reconstruction of the ensemble-averaged microbubble dynamics. Using simultaneous optical and acoustic measurements, we demonstrate recovery of microbubble oscillations with ~5% relative error for oscillation amplitudes up to ~15% of equilibrium radius. Finally, we show that oscillations within the framework's operating regime ($\Delta R/R_0 \leq 20\%$) generate sonoporation-relevant mechanical stress in vesicles as small as 10 μm (capillary number $Ca^c{}_K \geq 0.01$), under physiologically relevant conditions. Together, this work establishes a quantitative framework for acoustic emission–based monitoring of weakly nonlinear microbubble oscillations in clinically relevant, optically inaccessible environments to enable improved control of emerging ultrasound diagnostic and therapeutic technologies.

## Introduction

Microbubble (MB) contrast agents enable a broad range of diagnostic and therapeutic ultrasound (US) applications [1,2]. Historically, most US applications have exploited inertial cavitation and MB destruction, including imaging of contrast-agent replenishment [3] as well as sonoporation and mechanical ablation for therapy [4,5]. The prominence of MB destruction in US applications is, in part, due to the presence of coarse acoustic signatures, such as wideband emissions, that provide simple and robust indicators of cavitation activity and facilitate mechanistic interpretation of MB–tissue interactions [6]. By contrast, a growing class of US applications, including cell tracking [7] and modulation of the blood–brain barrier phenotype [8], depend on stable, nonlinear MB oscillations (i.e., non-inertial), where precise monitoring and control over the MB dynamics determines safety, efficiency, and biological responses. Likewise, increasing evidence suggests that the stress generated by stable microbubble oscillations plays a key role in balancing sonoporation efficiency and maintaining cell viability [5]. Quantitative characterization of these oscillations is therefore critical yet remains challenging because they occur on microsecond and micrometer scales.

Current quantitative methods rely almost exclusively on high-frame rate optical microscopy [9–13] that enables direct visualization of MB radius-time behavior. These optical techniques, however, are limited to *in vitro* settings [14] or superficial, optically transparent tissues [15,16], leaving MB behavior in deep or clinically relevant environments largely inaccessible. To address this limitation, acoustic monitoring of microbubble dynamics has emerged as a promising alternative. The most widely used strategy monitors MB acoustic emissions (AEs) (i.e., secondary pressure waves generated by bubble oscillations), using passive cavitation detectors (PCDs) [17]. Frequency-domain analysis of these emissions enables decomposition into harmonic, subharmonic, ultraharmonic, and wideband (i.e. non-harmonic or inharmonic) components, providing qualitative insights into oscillation mode and intensity (e.g., strong harmonic content without wideband emissions is generally indicative of stable oscillation). However, existing AE metrics that include mean AE level [18–20], cumulative dose [21], or conversion into physical units [22,23], do not provide quantitative information about the MB oscillation radius and thus serve only as indirect proxies for the underlying dynamics.

Recognizing this gap, a limited number of specialized acoustic methods have been proposed to recover physically meaningful MB dynamics. One such approach, termed "acoustical camera", reconstructs the absolute radius-time curve of freely floating bubbles by combining amplitude and phase measurements of high-frequency ultrasound pulses [24]. Although capable of resolving detailed radial motion, the technique relies on very high frequencies (i.e., above 10 MHz) and therefore suffers from significant acoustic attenuation, restricting its use to superficial targets. A second approach records MB acoustic signal using a calibrated transducer (e.g., PCD) to quantify the nanometer-scale radial dynamics using time-domain inversion of the recorded MB AEs [25]. Although this method can exceed the sensitivity of optical tracking under ideal conditions, because it is a time domain method it is inherently sensitive to noise and baseline offsets; hence to accurately recover MB dynamics, adequate filtering of higher order harmonics is required that may lead to loss of information. Moreover, this approach has not been extended beyond single-microbubble configurations, leaving a critical gap in the quantitative monitoring of microbubble populations under physiologically relevant conditions.

To address this unmet need, we introduce a framework, termed Linear Acoustic Wave Propagation and Superposition (LAWPS), for quantitatively estimating MB oscillation radius from passively recorded AEs. Leveraging Fourier series, a well-established method for characterizing nonlinear systems [26,27], we adapt this representation to the context of MB acoustics. We theoretically extend the classical monopole radiation theory (i.e., limited to single-frequency oscillations) to derive a relationship between AEs and weakly nonlinear MB radius dynamics containing multiple frequency components. Then, we further generalize the framework to the case of ensemble of microbubbles (**Fig. 1**). Finally, we experimentally validate LAWPS by comparing AE-derived radius–time curves to direct optical measurements obtained with high–framerate microscopy.

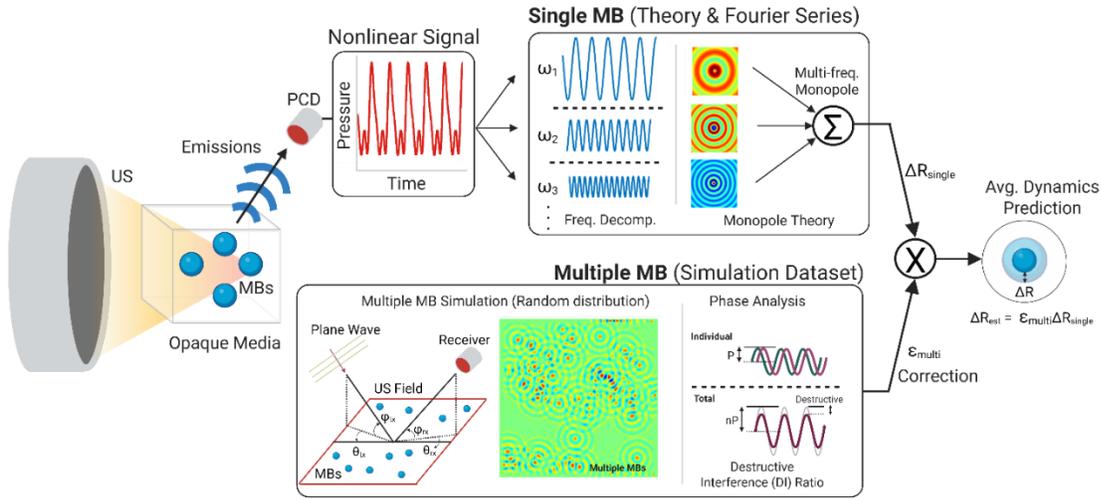

**Figure 1.** Flowchart for acoustic emission-based estimation of MB oscillation radius

## 1. Results

### 1.1. Derivation of relationship between single MB oscillation and AE

To examine the relationship between microbubble oscillation dynamics and the corresponding acoustic emission, we consider two simplifying assumptions: i) the oscillation of the MB is purely spherical and sinusoidal (i.e., weakly nonlinear) following the excitation pressure, and ii) MBs with an initial equilibrium radius $R_0$ behave as acoustic monopoles, such that $R_0 \ll \lambda$ and $\Delta R \ll \lambda$, with $\lambda$ being the acoustic wavelength. For a simple harmonic (linear) monopole oscillating at a single and fundamental angular frequency $\omega_0$, the bubble radius varies temporally as $R_{(t)}$, with a corresponding surface velocity $\dot{R}_{(t)}$ (Equation 1). Under steady-state excitation, the acoustic pressure field generated by such an oscillation can be obtained by combining Euler's equation (Equation 2) with the spherical wave solution (Equation 3), where r is distance from monopole to a point receiver.

$$\dot{R}_{(t)} = \dot{R} e^{j\omega_0 t} \tag{1}$$

$$\dot{R}_{(t)} = -\frac{1}{\rho_0} \int \nabla P_{(R_0, t)} dt \tag{2}$$

$$P_{(r,t)} = p_0 \frac{r_0}{r} e^{j(\omega_0 t - kr)} \tag{3}$$

By applying the boundary condition, where the particle velocity at the bubble surface ($r = R_0$) equals the bubble's surface velocity, we get the following general expression for the radiated pressure (Equation 4) [28,29], where $k_0 = \frac{\omega_0}{c}$.

$$P_{(r,t)} = \frac{\rho_0 \omega_0 R_0^2}{r} \frac{1}{\sqrt{(k_0 R_0)^2 + 1}} e^{j\{\arctan(1/k_0 R_0) - k_0(r - R_0)\}} \dot{R}_{(t)} \tag{4}$$

While this expression holds true for a linear (single) harmonic oscillator, MBs' dynamics are nonlinear and contain multiple frequency components. We solve this problem by decoupling the multi-frequency components of AE using Fourier Series Expansion (Equations 5 and 6), with *n* being an integer representing harmonics of the fundamental radial frequency $\omega_0$, which introduces nonlinearity to the monopole radiation theory; $\dot{R}_n$ and $P_n$ are complex Fourier coefficients.

$$P_{(r,t)} = \sum_{n=-\infty}^{\infty} P_n e^{nj\omega_0 t} \tag{5}$$

$$\dot{R}_{(t)} = \sum_{n=-\infty}^{\infty} \dot{R}_n e^{nj\omega_0 t} \tag{6}$$

Assuming that each frequency content in oscillation and its resulting AE is one-to-one correspondence (e.g., $2f_0$ oscillation → $2f_0$ AE, vice versa), we can establish a relationship between $n^{th}$ harmonic's complex Fourier coefficients of pressure and surface velocity (Equation 7 and 8) as follows:

$$P_n = \int_{-\frac{T_o}{2}}^{\frac{T_o}{2}} P_{(r,t)} e^{-nj\omega_0 t} \, dt = \frac{\rho_0 \omega_0 R_0^2}{r} \frac{n}{\sqrt{(nk_0 R_0)^2 + 1}} e^{j\{\arctan(1/nk_0 R_0) - nk_0(r-R_0)\}} \dot{R}_n \tag{7}$$

$$\dot{R}_n = \frac{r}{\rho_0 \omega_0 R_0^2} \frac{\sqrt{(nk_0 R_0)^2 + 1}}{n} e^{-j\{\arctan(1/nk_0 R_0) - nk_0(r-R_0)\}} P_n \tag{8}$$

Then, the original surface velocity can be calculated by summing all its harmonic frequency components (by substituting Equations 7 and 8 to Equations 5 and 6), as follows:

$$P_{(r,t)} = \sum_{n=-\infty}^{\infty} \frac{\rho_0 \omega_0 R_0^2}{r} \frac{n}{\sqrt{(nk_0 R_0)^2 + 1}} e^{j\{\arctan(1/nk_0 R_0) - nk_0(r-R_0)\}} \dot{R}_n \, e^{nj\omega_0 t} \tag{9}$$

$$\dot{R}_{(t)} = \sum_{n=-\infty}^{\infty} \frac{r}{\rho_0 \omega_0 R_0^2} \frac{\sqrt{(nk_0 R_0)^2 + 1}}{n} e^{-j\{\arctan(1/nk_0 R_0) - nk_0(r-R_0)\}} P_n e^{nj\omega_0 t} \tag{10}$$

Finally, the radius as a function of time (i.e., $R_{(t)}$) can be obtained by integrating the obtained surface velocity (Equation 11).

$$R_{(t)} = \int \dot{R}_{(t)} \, dt \tag{11}$$

The above formulation termed, Linear Acoustic Wave Propagation and Superposition (LAWPS) algorithm provides an analytical method to estimate the i) radiated pressure from the MB surface velocity (LAWPS propagation – Equation 9) or ii) MB surface velocity from recorded pressure (LAWPS backpropagation – Equation 10).

### 1.2. Validation of LAWPS propagation algorithm with single MB model

To validate the LAWPS propagation algorithm (Equation 9), we first generated a single pulsating MB (**Fig. 2 A, B**) using the MB mathematical model by Hoff et al. [30] that is given in Equation 12. The specific MB properties used can be found in Methods section. Briefly, we generated the MB model using 0.5 MHz frequency, excitation pressure ranging from 10 ~ 200 kPa, and MB sizes of $R_0 = 2 \sim 4$ μm. Note that the LAWPS algorithm utilizes the MB surface velocity and as such, it can be applied to any MBs with different properties and governing equations.

$$\rho_0 \left( \ddot{R}R + \frac{3}{2}\dot{R}^2 \right) = p_0 \left( \left(\frac{R_0}{R}\right)^{3\kappa} - 1 \right) - p_{acoustic}(t) - 4\mu_L \frac{\dot{R}}{R} - 12\mu_s \frac{d_{Se}R_0^2}{R^3} \frac{\dot{R}}{R}$$
$$- 12 G_s \frac{d_{Se}R_0^2}{R^3}\left(1 - \frac{R_0}{R}\right) \quad (12)$$

With the generated MB model, we compared and benchmarked the AE estimation by LAWPS propagation algorithm against an established AE estimation method proposed by Vokurka et al. [31,32] (Equation 13) (**Fig. 2 C**). For small amplitude and weak nonlinear oscillation ($\Delta R/R_0 < 20\%$), we found that AE estimation using LAWPS propagation algorithm agreed closely with Vokurka's (i.e., 5 % peak to peak error for n = 7, **Fig. 2 D**). Furthermore, contrary to Vokurka's estimation that assumes infinite speed of sound, the LAWPS algorithm incorporated a propagation delay (**Fig. 2 D**). Since we derived LAWPS based on frequency decomposition of AE (**Fig. 2D-E**), it allows choosing an order of approximation ("n" in Equation 9 & 10 – number of harmonics considered), similar to Taylor's series. We found that for LAWPS propagation, the order (i.e., number of harmonics considered) was proportional to accuracy of estimation (~ 5% mean error at 7$^{th}$ order, **Fig. 2 F**), which also converged faster for larger MB sizes. This can be intuitively observed in the term $\frac{n}{\sqrt{(nkR_0)^2+1}}$ in Equation 9; for monopole condition ($kR_0 \ll 1$), the numerator dominates, which suggests that higher oscillation frequencies are more efficient in radiating pressure under lossless conditions corroborating other studies on this matter [25].

$$AE = \rho \left[\frac{R}{r}(\ddot{R}R + 2\dot{R}^2) - \frac{1}{2}\left(\frac{R}{r}\right)^4 \dot{R}^2\right] \quad (13)$$

### 1.3. Validation of LAWPS backpropagation algorithm

As we mentioned above, the value of LAWPS algorithm lies in its theoretical potential to estimate the temporal oscillation radius change (R(t)) by backpropagating the recorded MB AE ($P_{(r,t)}$). To test it, we applied the LAWPS backpropagation algorithm (Equation 10) to Vokurka's AE estimation (**Fig. 2 D**, i.e., we applied LAWPS backpropagation to black curve). As before, we applied different orders for backpropagation, and our result suggested that 2$^{nd}$ order backpropagation was enough to reconstruct the original MB oscillation with less than 3% peak-to-peak error (**Fig. 2 G**). Interestingly, this indicates that pressure amplitudes of higher harmonic frequencies (> 3$^{rd}$ harmonic, or n > 3) have diminishing effects in the MB oscillation reconstruction (see also Equation 10; the term $\frac{\sqrt{(nkR_0)^2+1}}{n}$, which decays as a factor of 1/n for $kR_0 \ll 1$). Consequently, the LAWPS algorithm can be used with narrow bandwidth PCD, provided that the receiver is able to capture at least up to 2$^{nd}$ harmonic. Moreover, in experimental conditions, higher order harmonic (higher frequency) waves are more prone to propagation-related losses (e.g., scattering and absorption) which hinder their reliable estimation. The reduced dependency on these higher order components therefore favors the practical application of LAWPS backpropagation algorithm (**Fig. 2 H-I**).

Finally, we assessed the potential limitations of LAWPS algorithm. To do so, we generated AE using Vokurka's formulation for stronger MB oscillations ($\Delta R/R_0 = 20\% \sim 40\%$) and applied LAWPS backpropagation with varying $R_0$ values. Our results revealed that MB oscillation ($\Delta R/R_0$) greater than 30% (close to rupturing [12]) generated sharp transient responses in AE (i.e., inertial response), and for such oscillations (i.e., outside of weakly nonlinear regime) the LAWPS backpropagation resulted in larger errors (**Fig. S1 A**). While sub-harmonic-based multiples in LAWPS allow incorporation of sub- and ultra-harmonics, they still cannot effectively capture inertial responses, which underscore the fundamental limitation of Fourier-series based reconstruction under high amplitude inertial oscillations (**Fig. S1**). We also found that LAWPS backpropagation is sensitive to $R_0$. Using an incorrect $R_0$ (e.g., 30% error from correct $R_0$) resulted in a relative backpropagation error of up to 15% (**Fig. 2 J-K**), which highlights the importance of accurate prior knowledge of the MB initial radius. In aggregate, our results demonstrate that for small oscillations ($\Delta R/R_0 < 20\%$) of a MB with known radius the LAWPS algorithm can effectively convert the AE (i.e. backpropagate) into MB oscillation dynamics.

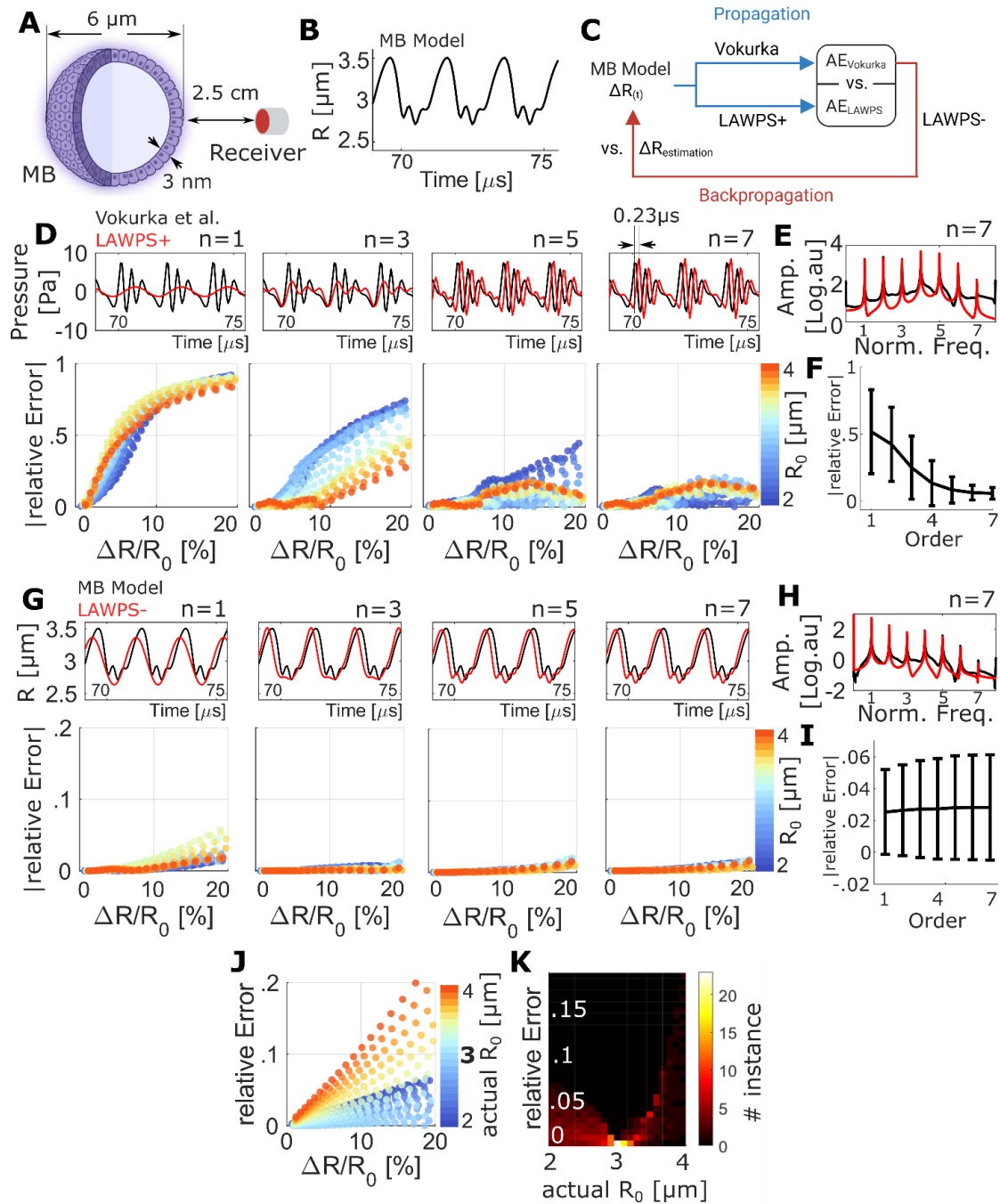

**Figure 2. Single MB radius estimation. A)** Microbubble model geometry: 6 μm diameter with 3 nm thickness (e.g., lipid shell) with the receiver placed at 2.5 cm from the microbubble. **B)** MB mathematical

model: The excitation and physical parameters are provided in Methods. **C)** LAWPS validation methodology. AE calculated from MB model using LAWPS propagation (LAWPS+) versus Vokurka's method was compared. Then, estimation of MB model was calculated using LAWPS backpropagation (LAWPS-) onto Vokurka's estimation and compared with original model. **D)** AE estimation at different orders. Top: LAWPS propagation estimation (red) and Vokurka's estimation (black). Two algorithms have phase difference of 0.23 µs from 2.5 cm propagation. Bottom: Effect of different LAWPS+ order, initial radius, and oscillation strength onto absolute relative error of AE propagation. **E)** Frequency spectrum of LAWPS algorithm and Vokurka's method. **F)** Effect of order (n) and oscillation strength onto LAWPS propagation error. **G)** LAWPS backpropagation algorithm applied onto Vokurka's AE estimation. Top: LAWPS reconstruction (red) compared to original MB model (black). Bottom: Effect of different LAWPS- order, initial radius, and oscillation strength onto absolute relative error of MB radius change reconstruction. **H)** Frequency spectrum of LAWPS algorithm and original MB mathematical model. **I)** Effect of order (n) and oscillation strength onto LAWPS backpropagation error. **J, K)** Effect of initial radius input in backpropagation onto backpropagation error. $R_0 = 3$ µm was used for all backpropagations, which were in fact from different actual $R_0$ (actual $R_0 = 2\sim 4$ µm.)

## 1.4. Multiple MB scheme and phase-based correction

While promising under small oscillation conditions, extension of the LAWPS single MB algorithm to multiple distributed MBs is important for applying it under realistic conditions. To do so we must account for the impact of i) the secondary pressure generated by adjacent MBs on MB dynamics (i.e., MB-MB interaction), and ii) destructive interference on the recorded emissions (i.e., propagated secondary pressure) from the distributed sources. To investigate the effect of MB-MB interaction in ensemble of MBs, we first simulated a MB cluster where monodispersed MBs of known radius are randomly distributed in a 2D field with steady state, homogeneous, and in-phase MB dynamics (mono-acoustic behavior). Based on a modification of Vokurka's AE model extended to multiple MBs scheme and with propagation phase delay, as shown below.

$$AE = \sum_{a=1, a\neq b}^{N-1} \left[\rho \frac{R}{r_{ab}}(\ddot{R}R + 2\dot{R}^2) - \rho \frac{1}{2}\left(\frac{R}{r_{ab}}\right)^4 \dot{R}^2\right] e^{jk(r_{ab}-R_0)} \qquad (14)$$

We then propagated secondary pressure from neighboring MBs to a MB located at center of field to be compared to the excitation pressure. We repeated the simulation for varying concentration of MBs (void fraction $\beta = 0.05 \sim 0.5$ %, expected MB-MB distance of $10 \sim 30\ R_0$, see Methods section and **Table 2** for calculation) in the US field and varying excitation pressure ($30 \sim 80$ kPa). For each of those conditions, we simulated 120 random MB clusters (Poisson disk sampling, see Methods). We found that for void fraction of $0.05 \sim 0.1$ %, each MBs experienced secondary pressure (peak-to-peak) from neighboring MBs that was less than approximately 4% of the excitation pressure (**Fig. 3 A**). For higher concentration of MBs ($\beta = 0.5$ %), each MBs experienced secondary pressure close to 20% of primary excitation pressure (**Fig. 3 B**). We then generated a MB model using the affected pressure amplitude (i.e., initial excitation and secondary contribution from multiple MBs) as input. For low MB concentrations ($\beta = 0.05 \sim 0.26$%: mean MB-MB distance greater than 14 $R_0$, **Table 2**) and oscillation amplitudes $\Delta R/R_0 < 20$%, our analysis indicates less than 2% mean peak-to-peak change from the original oscillation (i.e., prior to secondary effects, **Fig. 3 C-E**). While increasing number of MB results in higher pressure contribution from neighboring MBs, our findings suggest that in low concentration conditions shown above, the excitation pressure is still the major factor in inducing MB dynamics compared to the secondary MB-MB acoustical interactions (**Fig. 3 E**).

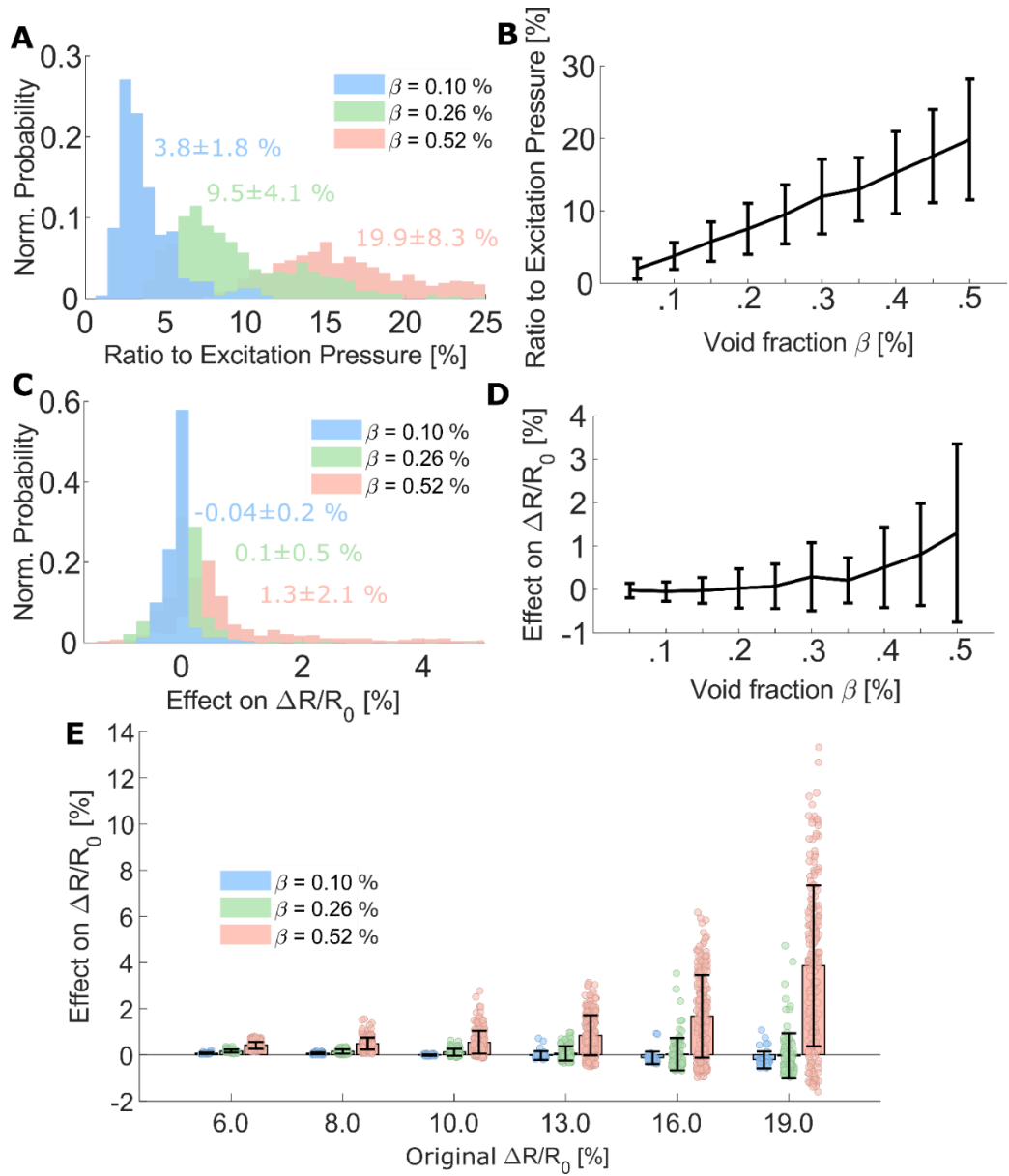

**Figure 3. Multiple MB acoustic interaction. A** and **B)** show the effect of different MB concentrations (β = 0.05 ~ 0.5 %) onto secondary pressure applied on MB at center of field. **C** and **D)** shows the resulting MB oscillation radius change after the secondary pressure effects for different MB concentrations (β = 0.05 ~ 0.5 %). **E)** Effect of original MB radius change onto MB radius change after MB-MB interaction. Error bars indicate standard deviation. See **Table 2** for calculation of void fraction β and expected MB-MB distance.

Multiple MBs scheme is additionally challenging due to the uncertainties coming from spatial distribution of the MBs. In realistic settings, MBs rarely exist in isolation, and their spatial variability leads to complex phase delays that can mask true oscillatory dynamics (**Fig. 4 A**). Specifically, depending on the source/receiver location, MBs in the US field experience 1) excitation delays from the source and 2) receive delays to the receiver. To study the effect of MB spatial distribution onto propagated pressure, we simulated randomly distributed MBs in US field: 1000 distributions for each of varying number of MBs (i.e., concentration, ranging from β = 0.11 ~ 0.56%, see Methods for calculation), source/receiver geometries, and oscillation frequencies (**Fig. 4 B**). We then quantified the onset of pressure propagated to the receiver and compared it to an ideal constructive interfered pressure (i.e., when all MBs are located at a single point in the center of the field) (**Fig. 4 C**) and defined such ratio as destructive to constructive interference ratio (DI/CI ratio). Interestingly, we found that the amount of destructive interference for 1000 random spatially distributed samples (i.e., MB cluster samples) formed a normal distribution (**Fig. 4 D**) with its mean independent from number or concentration ($\beta = 0.11 \sim 0.56\%$) of MBs but its deviation inversely proportional to concentration of MBs in the field (**Fig. 4 E**). This suggests that there exists a representative DI/CI ratio that reflects a specific MB distribution geometry (e.g., US field shape such as vessels; source/receiver allocation). The mean of the distribution was very sensitive to the source/receiver geometry (**Fig. 4 F**) and monopole's oscillation frequency (**Fig. 4 G**). However, our findings suggest that they can be accounted for, as the DI/CI ratio formed a distribution that can be represented with a mean value. Crucially, our analysis revealed that when the source and receiver are aligned in a line facing each other, the effect from destructive interference is minimized (DI/CI nearly 100%, **Fig. 4 H**). Our multi-MB simulation scheme revealed a similar predictable behavior (i.e. statistical pattern) in US frequency (**Fig.4 H**). This behavior allows the application of LAWPS onto multiple MB scheme, as the DI/CI ratio can be used to correct the impact of destructive interference to the data, for example, by dividing the recorded signal by DI/CI ratio (e.g., **Fig. 4 A**, estimate left from right). Moreover, it revealed optimal source and receiver configurations that can be used for calibrating and designing clinical systems.

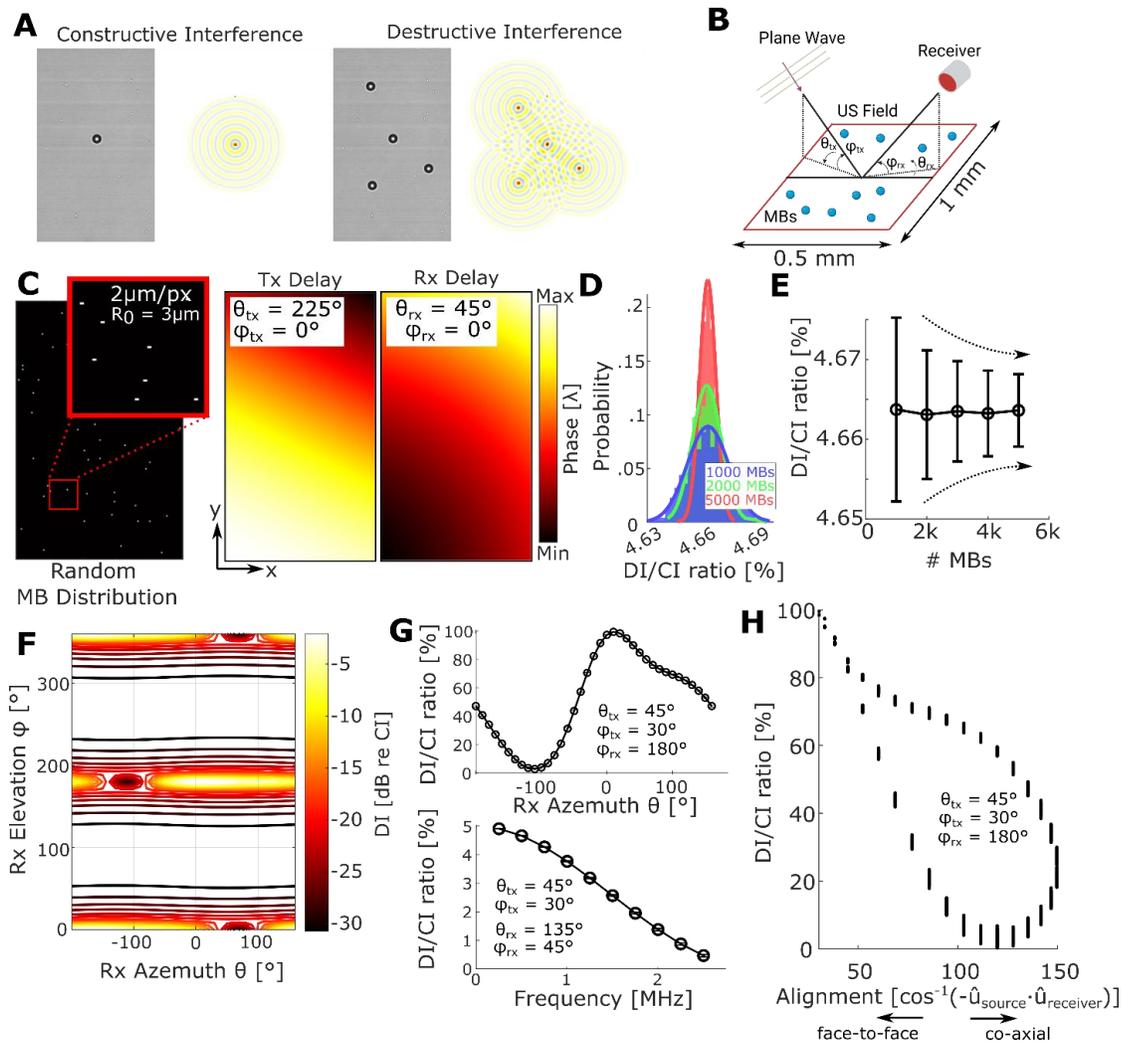

**Figure 4. Multiple MB correction using simulated MB distribution and statistical inference.**
**A)** A representative image of MBs spatial distribution contributing to destructive interference. **B)** Multi-monopole scheme simulation setup. Variables include source/receiver geometry (elevation φ and azimuth θ. rx = receiver, tx = plane wave source), frequency, and number of MBs in the field. **C)** Example random MB distribution used in simulation. Poisson disk sampling method was used to generate random MB ensemble (left) and delays from excitation (middle) and reception (right) were calculated for each variation to be applied onto the ensemble. **D)** and **E)** Effect of number of MBs (1000~5000 MBs; equivalent to β = 0.11 ~ 0.56%) in destructive to constructive interference ratio (DI/CI ratio). Mean of DI/CI ratio was independent of the number of MBs but the variance of DI/CI ratio distribution was inversely proportional to the number of MBs. Source/receiver orientation of $\theta_{tx} = 45°$, $\theta_{rx} = 135°$, $\varphi_{tx} = 30°$, $\varphi_{rx} = 45°$. **F)** Effect of source/receiver geometry onto DI/CI ratio seen in 2D. There exists an optimal source/receiver alignment (face-to-face alignment) that results in minimal destructive interference. **G)** Effect of source/receiver geometry onto DI/CI ratio seen in 1D (top) and effect of monopole frequency onto DI/CI ratio at source/receiver orientation of $\theta_{tx} = 45°$, $\theta_{rx} = 135°$, $\varphi_{tx} = 30°$, $\varphi_{rx} = 45°$ (bottom). Error bars indicate standard deviation (i.e., smaller than the data). **H)** Facing-angle error derived from source/receiver orientation versus DI/CI ratio. A value of 0° alignment indicates source and receiver is facing each other, whereas larger errors indicate increased misalignment (i.e., coaxial alignment)

## 1.5. Experimental validation of LAWPS algorithm using concurrent acoustic and optical monitoring of MB dynamics

To assess the LAWPS algorithm experimentally, we designed a setup integrating a high frame rate microscopy (5 Mfps) with a precisely aligned and calibrated acoustic system consisting of a 0.5 MHz source and a PCD (**Fig. 5 A**, see Methods for their calibration – **Fig. S2**). The platform was built around a 200 μm height microfluidic channel (μ-slide I Luer, ibidi, Germany). Interestingly, our k-wave simulation of the ibidi μ-slide using multiple monopoles indicated that the mean AE loss through the channel was approximately 6% compared to free field, possibly due to the combined effect of reflection/transmission from ibidi slide and monopole's behavior next to hard/soft boundaries (**Fig. S3**). To reduce uncertainty from the initial MB radius on the estimated MB radius oscillation, we employed MBs (BG9732, Bracco) with average size distribution of 6 ± 0.43μm, and polydispersity index of 7% (**Fig. 5 B**). To minimize MB-MB acoustic interactions, as informed by our prior simulation results (**Fig. 3 E**), we conservatively diluted the MBs 100 times – corresponding to a volume fraction (β) of 0.015 % based on manufacturer's specifications (see Methods) – and also optically confirmed the average MB-MB spacing to be greater than 15 $R_0$ before sonication. Meanwhile to confirm a minimal potential translation motion (**Fig. S4**), which may violate our monopole assumption, we acquired post-sonication images after each experiment.

Using varying excitation pressure ranging from 100 to 250 kPa, we recorded acoustic and optical data simultaneously after waiting for MB to float to ibidi surface (**Fig. 5 C**, see Methods section for detailed MB administration protocol during high framerate microscopy). To compensate for destructive interference in this setup ($\theta_{tx} = 45°$, $\theta_{rx} = 135°$, $\varphi_{tx} = 30°$, $\varphi_{rx} = 45°$), we calculated and used a DI/CI ratio of 4.67% (**Fig. 5 D**). For other harmonic frequencies, we referred to our simulation for frequency specific DI/CI ratios (**Fig. 4 G**, 1 MHz – 3.76%, 1.5 MHz – 2.56%, 2 MHz – 1.37%, 2.5 MHz – 0.04%). Finally, to estimate the MBs radius change, we first isolated the MB signal by subtracting a carefully aligned background signal in time domain (**Fig. 5 E**). Then, we applied 5$^{th}$ order (n = 5$^{th}$ order corresponds to Nyquist frequency of our high framerate microscopy – 5 Mfps) LAWPS backpropagation onto the acquired US signal while compensating for the ibidi channel's reflection/transmission, the DI/CI ratio of current setup (**Fig. 5 F**). We assumed $R_0 = 3$μm (with ± 0.43μm tolerance range to account for small polydispersity – upper/lower estimations shown in red; **Fig. 5H**) and determined the number of MBs from the optical images. We compared the results with images acquired using high framerate microscopy (**Fig. 5 G**). We found that upon such corrections peak-to-peak oscillation estimated by LAWPS algorithm agreed closely with the high frame rate camera recordings while also capturing MBs' harmonic oscillations (**Fig. 5 H**). While we observed sub-harmonic components in some of recorded AEs, it was not optically observed in MB oscillations, possibly due to surface modes [25] which were not capturable due to our optical resolution limits (**Fig. S5**). However, as expected, our result shows that the accuracy of LAWPS is the highest at weak MB oscillation having relative oscillation radius of $\Delta R/R_0 < 20\%$ (**Fig. 5 I**). Together our findings demonstrate the potential of LAWPS algorithm to quantify the weakly nonlinear MB dynamics in solutions with multiple monodisperse MBs.

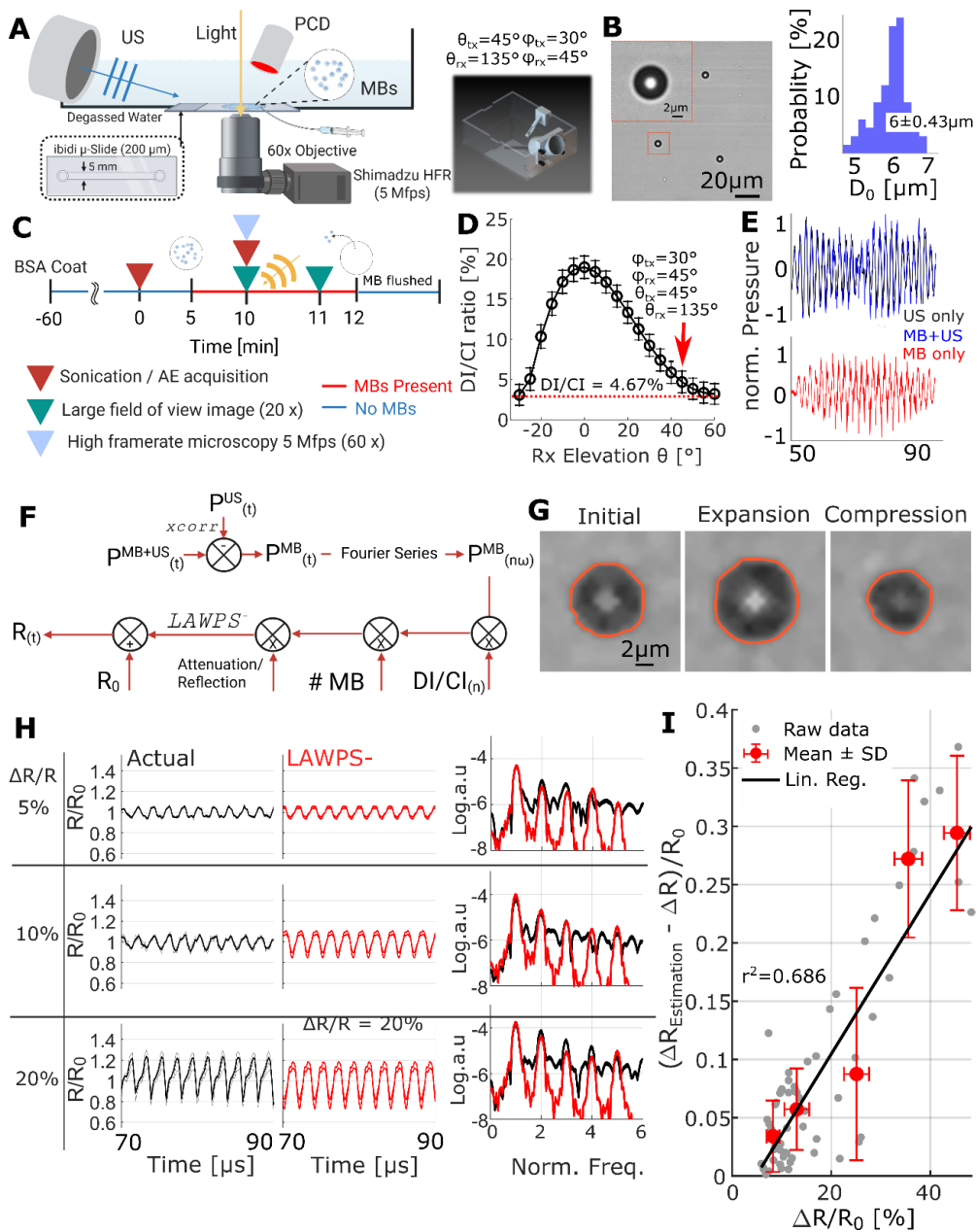

**Figure 5. Estimation of MB radius dynamics in in vitro setup using LAWPS. A)** A representative in vitro validation setup. High framerate microscopy was coupled with PCD. **B)** Monodisperse MBs (BG9732 6μm, Bracco) used for in vitro validation. **C)** Timeline of data acquisition. Both acoustic and optical data were precisely timed and simultaneously acquired. Background AE was recorded prior to MB injection, and large field of view image was acquired before sonication. **D)** Calculated 4.67% DI/CI ratio for current in vitro setup at 0.5 MHz. For other harmonic frequencies, we referred to **Fig. 4G** (1 MHz – 3.76%, 1.5 MHz – 2.56%, 2 MHz – 1.37%, 2.5 MHz – 0.04%). **E)** Representative figure showing background signal removal. AE was recorded before MB injection and was used for subtraction from MB AE after MB injection. **F)** A protocol for in vitro LAWPS backpropagation implementation. **G)** Representative images from high framerate microscopy showing initial radius/expansion/compression at small oscillation under 200 kPa pressure. The orange lines show the MB edges for clarity. **H)** Comparison of LAWPS backpropagated estimation of oscillatory radius versus actual optical measurement. Left: Actual (optical observations) measurement of MB radius change (gray indicate different MBs and black shows average radius change in same field of view), Middle: 5th order LAWPS reconstruction (upper/lower lines represent confidence intervals from ± 0.43μm polydispersity, Right: frequency spectra of actual measurement (black) and LAWPS reconstruction (red). For 5% and 10% oscillations batches, we used excitation pressure of 100 kPa. For 20% oscillation, we used 115 kPa. **I)** Error of LAWPS estimation under various oscillatory strength with linear regression (black). n = 69 MBs (gray). Red error bars indicate mean and standard deviation.

### 1.6. Sonoporation-relevant mechanical effects emerge within the LAWPS operating regime

To assess whether the operation regime of LAWPS algorithm ($\Delta R/R_0 < 20\%$) is biologically relevant, we theoretically evaluated the resulting mechanical stress. We considered a MB with $R_0$ = 3 μm radius oscillating near a rigid boundary (e.g., vessel wall) and first evaluated the bubble-streaming Reynolds number $Re_{bs}$ [33,34] using Equation 15, where $\nu$ is the kinematic viscosity of the surrounding medium. For $\Delta R/R_0 = 10\%$ (an oscillation regime with ~5% LAWPS backpropagation accuracy), $f_0$ = 0.5 MHz, and $\nu = 1.0 \times 10^{-6} m^2 \cdot s^{-1}$, we obtained $Re_{bs} \approx 0.05$. This satisfies $Re_{bs} < 1$, validating the use of Rayleigh-Nyborg-Westervelt (RNW) streaming and the Stokes flow approximation to characterize the induced microstreaming [35].

$$Re_{bs} = \left(\frac{\Delta R}{R_0}\right)^2 \left(\frac{2\pi f_0 R_0^2}{\nu}\right)^{1/2} \quad (15)$$

Under these conditions, the subsequent maximum tangential streaming velocity near the cell was estimated as $u \approx 2\pi f_0 (\Delta R/R_0)^2 R_0 \approx 94 \; mm \cdot s^{-1}$ [34]. Moreover, the maximum strain rate, $G_{max}$, can be approximated in the small-oscillation regime – where radial and translational oscillations are comparable – using Equation 16. For $\Delta R/R_0 = 10\%$, we obtained $G_{max} \approx 31.4 \times 10^3 s^{-1}$.

$$G_{max} \approx 2\pi f_0 \left(\frac{\Delta R}{R_0}\right)^2 \quad (16)$$

At a given strain rate, the resulting maximum shear stress ($S_{max}$) and the associated capillary number $Ca_K$ depend on the dynamic viscosity $\eta$ of the medium, the characteristic vesicle (cell) radius D [34,36,37], and area expansion modulus $K_A$ (for DOPC, 0.24 Nm$^{-1}$) as described by Equation 17.

$$Ca_K = \frac{S_{max}}{K_A} = \frac{\eta G_{max} D}{K_A} \quad (17)$$

Using a critical capillary number $Ca^c_K \approx 0.01$, above which reversible membrane rupture is expected to occur [36], we delineated the theoretical cell-membrane-rupture regime associated with LAWPS-compatible oscillations (**Fig. 6**). In water-like media ($\eta = 1.0 \; mPa \cdot s$), conditions satisfying both low $Re_{bs}$ (Stoke's flow) and membrane rupture were limited to larger vesicles and stronger oscillations. In contrast, for media with dynamic viscosity greater than 6 times that of water (i.e. similar to blood), oscillations well within the LAWPS operating regime were sufficient to generate capillary numbers indicative of membrane rupture for vesicle radii as small as D = 10 μm. Collectively, our results suggest that LAWPS operating regime encompasses oscillation strengths capable of producing sonoporation-relevant mechanical stresses for physiologically relevant cell sizes, despite remaining within the weakly nonlinear and stable oscillation domain.

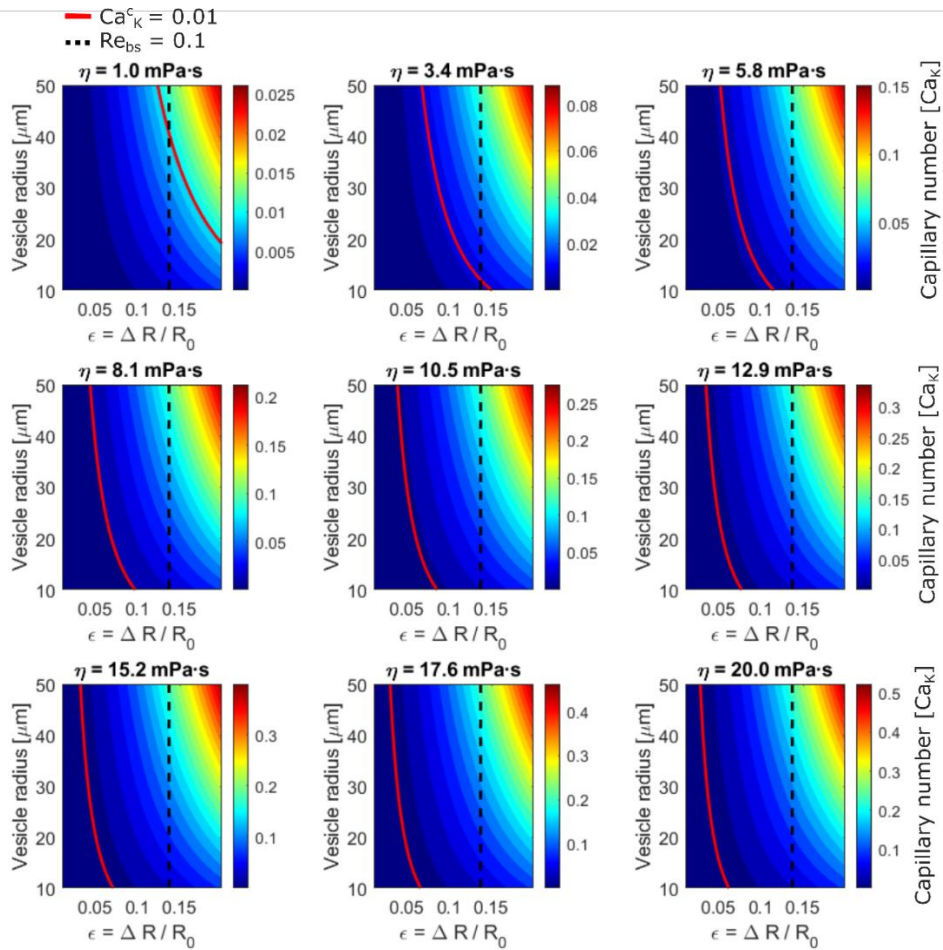

**Figure 6. Sonoporation-relevant capillary numbers are within the LAWPS operating regime.**
**A)** Parametric maps of the capillary number $Ca_K$ as a function of normalized MB oscillation amplitude $\varepsilon = \Delta R/R_0$ and vesicle (cell) radius D, evaluated for increasing medium dynamic viscosity $\eta$ (1~20mPa.s). Color contours indicate the magnitude of $Ca_K$. The red solid curve denotes the critical capillary number for reversible membrane rupture $Ca^c_K = 0.01$, whie the black dashed line marks the bubble-streaming Reynolds number threshold ($Re_{bs} = 0.1 < 1$), below which Stokes-flow and RNW streaming assumptions remain valid. As medium viscosity increases, oscillation amplitudes well within the LAWPS operating regime $\Delta R/R_0 < 20\%$ become sufficient to exceed the membrane rupture threshold for physiologically relevant cell sizes, indicating that LAWPS-compatible MB oscillations can generate sonoporation-relevant mechanical stresses despite remaining in the weakly nonlinear regime.

## 2. Discussion

In this study, we introduced and validated a linear acoustic wave propagation and superposition - LAWPS - framework for estimating the oscillatory radius of a monodisperse MB population directly from their acoustic emissions (AE). Our theoretical derivation extends the classical linear monopole radiation theory to weakly nonlinear (multiple frequency) oscillations via harmonic decomposition from Fourier series; conceptually, the LAWPS framework can be viewed as analogous to a Taylor series expansion for MB dynamics, capturing the dominant linear and low-order harmonic terms essential in describing oscillatory behaviors (**Fig. 5 H**). While back-propagation approaches have been previously explored, they have largely been focused on imaging and source localization applications [38] or have relied on single-MB time domain-analysis [25]. In contrast, LAWPS uniquely leverages the frequency-domain backpropagation to quantitatively recover MB oscillatory dynamics from multi-harmonic AE measurements. Furthermore, by incorporating MB population-based correction factors, LAWPS bridges the gap in the quantitative monitoring of microbubble ensembles that represent realistic conditions. With its demonstrated theoretical accuracy under relatively small and weakly nonlinear oscillation regime, our method holds potential for MB dynamics monitoring.

Our theoretical analysis allowed us to clarify the relationship between MB oscillation and AE. First, we demonstrated that AE is proportional to oscillation frequency and MB size. This corroborates prior single-MB study that utilized linearized time-domain method, which suggested $AE \propto \omega^2$ and $AE \propto R_0^3$ [25]. Due to this proportional relationship, estimating AE from single oscillating MB is highly susceptible to signal-to-noise ratio where noise can result in broadband emission that is amplified with such relationship, resulting in errors in the estimation of MB dynamics. In contrast, for backpropagation of relatively small oscillations ($\Delta R/R_0 \leq 20\%$), contributions of higher-order harmonics are less critical (Equation 10), which is consistent with previous findings [25]. This indicates that LAWPS-based radius estimation does not require high-order harmonic measurements. Higher-frequency harmonics are more strongly affected by attenuation (scattering and absorption), making them less reliable experimentally. By focusing on selected frequency bands, the LAWPS backpropagation improves robustness and suitability for experimental and *in vivo* use.

Our analysis also revealed that specific source-receiver geometries (i.e., orientation and angle of incident) can minimize destructive interference due to phase cancellation; this optimal condition occurs when the source and receiver are directly facing one another. Crucially, our study highlights a configuration that is in conflict with currently used PCD geometries that employ coaxial alignment rather than face-to-face placement [18,39]. These findings provide new design considerations for improving sensitivity of AE-based MB dynamics monitoring and system calibration. Moreover, compared to conventional PCD methods that primarily quantify AE amplitude or energy to infer exposure strength [18,19,21], our approach establishes an explicit physical relationship between emitted pressure and MB surface velocity. This utilization of AE provides a potential pathway to extend MB dynamics monitoring efforts that are primarily focused on the type and relative strength of the oscillation to detailed information about the oscillation radius. As our LAWPS can be applied to optically opaque media, it has the potential to be useful under *in vivo* conditions in regimes relevant to stable cavitation, which can be further augmented by reducing receive errors using more sensitive and broadband hardware such as CMUT technology [40–42]. For applications that exploit inertial cavitation and MB destruction, this analysis can be used to calibrate the exposure settings, before scaling it up to the inertial cavitation regime.

Importantly, the microbubble oscillations within the framework's operating regime ($\Delta R/R_0 \leq 20\%$) can exert sonoporation-relevant mechanical stress. Under this regime, our theoretical calculations indicated strain rate of $G_{max} \approx 31.4 \times 10^3 s^{-1}$, which is comparable to previous findings with different MB conditions and excitation frequency ($G_{max} \approx 35.3 \times 10^3 s^{-1}$ for $R_0$ = 1.25 μm and $\Delta R/R_0$ = 7% [34] and $G_{max} \approx 10^3 s^{-1}$ for $R_0$ = 20 μm and $\Delta R/R_0$ = 5% [33]). While under water-like medium, such strain rate is not able to produce notable cell membrane rupture [33,34,36,37], at fluid with higher viscosity (i.e., similar to blood) the capillary number can reach membrane rupture level for physiologically relevant cell sizes [36]. This further highlights the potential of LAWPS algorithm infer acoustic streaming amplitudes [35,43] pertinent to a range of interventions such as sonoporation and possibly modulation of blood–brain barrier phenotype.

Our approach has a few limitations. First, our model assumes point receivers, which deviate from realistic detector geometries. However, we anticipate that the effect coming from finite aperture largely cancels out in the calculation of DI/CI ratio, since both destructive and constructive interference conditions were simulated under the same point-receiver assumption. Moreover, as indicated by the model and experimental validation (using high framerate optical microscopy) the accuracy is the highest when the MBs diameter is known; the accuracy relies on assumption that monodisperse MBs exhibit mono-acoustic behavior in ensembles. These findings emphasize the importance of using monodisperse MBs in experimental and therapeutic contexts to maximize predictive accuracy. In the strong oscillation regime (>30% radius change), LAWPS prediction accuracy degraded, reflecting the algorithm's assumption about purely spherical oscillations (i.e., weakly nonlinear). While this represents a limitation, it also delineates the operating window in which the framework is most reliable. For LAWPS to be applied under *in vivo* conditions, the formulation needs to be extended to account for scattering and distribution of MB (i.e., the shape of MB distribution such as vessel structure). Such extension could benefit by integrating thorough simulation for *in vivo* settings to quantify the acoustic loss; moreover, MBs' *in vivo* spatial distribution (which differs from our multiple-MB simulation's rectangular geometry) may be captured using knowledge of the underlying vascular distribution. This could potentially be captured using ultrasound localization microscopy [44].

Moreover, although our study employed low MB concentrations, we still observed measurable MB centroid spatial shifts following sonication (**Fig. S4**), likely arising from primary radiation forces and possibly secondary Bjerknes interactions. These displacements were minimal compared to the wavelength (3mm), confirming that the MBs behaved as point sources. Nevertheless, accounting for high MB concentration that may promote secondary Bjerknes forces and changes in MB dynamics (e.g., coalescence) might be needed [45]. Likewise, more detailed assessment of the impact of vessel-wall on MB dynamics and its potential in inducing non-spherical oscillations is required to extend the applicability of LAWPS to *in vivo* conditions. Despite these limitations, our algorithm is expected to remain both accurate and computationally efficient, provided that the assumptions of small-amplitude spherical oscillation and low-concentration conditions along with the proposed mitigation strategies (e.g., DI/CI ratio, number and spatial distribution of MBs) are applicable.

Collectively, our results establish LAWPS as a strategy to bridge the current gap between AE-based monitoring in optically opaque media (at least 30 mm depth) and direct observation of MB dynamics. By enabling quantitative, noninvasive estimation of MB oscillation radius, LAWPS could provide a robust indicator of cavitation activity and sonoporation-relevant mechanical stress and facilitate mechanistic interpretation of MB–tissue interactions. This may enhance the safety and efficacy of MB-FUS therapies by allowing real-time feedback control that ensures oscillations remain within desired regimes. Moreover, the simplicity of our framework also carries potential to more hardware-efficient platforms such as circuit-level accelerators or embedded processors [46–48], which pave a way for real-time integration to US systems. Overall, our investigations established LAWPS as a quantitative, noninvasive framework for monitoring microbubble dynamics and provided a theoretical framework for acoustic metrology, cavitation physics, sonochemistry, and ultrasound-based control of soft matter and biological systems.

## 3. Experimental Section

*Microbubble model with shell* – To validate the derived LAWPS algorithm, we generated MB oscillation using the model by Hoff et al. (Equation 12) with the properties provided in Table 1.

**Table 1. MB mathematical model properties**

| Property | $\rho_0$ | $p_0$ | $\mu_L$ | $\mu_S$ | $d_{Se}$ | $G_S$ |
|---|---|---|---|---|---|---|
| Value | 998 | 100 | 0.001 | 1.2 | 3 | 60 |
| Unit | kg/m$^3$ | kPa | Pa.s | Pa.s | nm | MPa |

*Theoretical validation of LAWPS algorithm* – To validate LAWPS propagation algorithm, we first utilized the Hoff's MB model ($P_{acoustic}$ = 10 ~ 200 kPa, $R_0$ = 6 µm) and applied Vokurka's AE estimation method (Equation 13). Then, we compared Vokurka's estimation with LAWPS propagation algorithm (Equation 9), with the varying order of approximation (n = 1~5). We then applied the LAWPS backpropagation onto the AE estimated by Vokurka's method and compared the resulting time series R with original MB model. The relative errors were computed by comparing the peak to peak of each compared method. A US Patent has been filed on aspects of this algorithm [49].

*Generation of MB cluster (Poisson disk sampling)* – To randomly distribute MB centers with a minimum separation distance $2R_0$, we employed Poisson disk sampling [50], which is a stochastic process that generates points within a bounded domain $\Omega \subset \mathbb{R}^2$ such that no two points lie closer than $2R_0$. Formally, the set of the generated points (i.e., MB locations with n MBs) $M = \{x_i \in \Omega\}_{i=1}^{n}$ satisfies:

$$\forall x_i, x_j \in M, i \neq j \text{ and } \|x_j - x_j\| \geq 2R_0$$

Our Poisson disk sampling algorithms proceeds iteratively by randomly proposing new candidate points within the domain and accepting them only if they satisfy the distance constraint ($2R_0$) relative to all previously placed points.

*Acoustic MB-MB interaction* – To investigate Acoustic interaction of MBs we first generated random location of MBs using Poisson disk method (10 ~ 100 MBs per 600 μm x 600 μm US field, void fraction β = 0.05 ~ 0.5 %), whose void fraction is given by Equation 18 [51], assuming channel height of $2R_0$, where N is number of MBs per unit volume.

$$\beta = \frac{4}{3}\pi R_0^3 N \tag{18}$$

For further reference, we also calculated the expected MB-MB distance using Equation 19 in 2D space [52], where N is number of MBs in unit area. Calculated void fraction and expected MB-MB distance are shown in **Table 2** for various concentrations used in the simulation.

$$Expected\ Distance = \frac{1}{2R_0\sqrt{N}} \tag{19}$$

Then, assuming mono-acoustic behavior of MBs, we applied varying excitation pressure to Hoff's model (30 ~ 80 kPa) and propagated the resulting AE, with the propagation delay incorporated (Equation 14). Then, we gathered the combined AE at the center of the field and analyzed how much it affects the original MB dynamics by applying the resulting amplitude change onto the Hoff's model.

**Table 2. MB concentration calculation**

| # MB in 600 μm x 600 μm US field | 10 | 20 | 30 | 40 | 50 | 60 | 70 | 80 | 90 | 100 |
|---|---|---|---|---|---|---|---|---|---|---|
| β [%] | 0.05 | 0.1 | 0.16 | 0.21 | 0.26 | 0.31 | 0.37 | 0.42 | 0.47 | 0.52 |
| MB-MB Distance | 31.6$R_0$ | 22.3$R_0$ | 18.3$R_0$ | 15.8$R_0$ | 14.1$R_0$ | 12.9$R_0$ | 12.0$R_0$ | 11.2$R_0$ | 10.5$R_0$ | 10.0$R_0$ |

*Multiple MBs DI/CI ratio* – To investigate AE loss due to MB spatial distribution, we generated random location (in 1 mm x 0.5 mm 2D field, assuming 0.2 mm channel height of ibidi slide) of MBs using Poisson disk method (with minimum 6 μm apart). Then by varying 1) the number of MBs in the field 1000 - 5000 MBs, which is equivalent to β = 0.11 ~ 0.56%, 2) the spatial location of source (0~180 degrees azimuth and altitude) spatial location of receiver (0~180 degrees azimuth and altitude), and 3) the emitted AE frequency (0.5 ~ 3.5 MHz), we simulated 1000 different MB cluster samples of each condition combination. At each frequency, the propagated AEs from each MBs (mono-acoustic / monodisperse monopoles at steady state) were assumed to be single frequency. Then, the resulting AEs were added at point receiver's location. This resulting pressure was then compared to ideal pressure from a "constructive interference case", where all monopoles were assumed to be in the center of the field. We defined the destructive interference to constructive interference ratio (DI/CI) and gathered DI/CI ratio for 1000 samples for each combination of variables for further analysis. Then, we found the mean DI/CI ratios for each condition to represent such combination. While using a point receiver may not represent the realistic PCD, the ratio operation (dividing DI case with CI) results in cancellation of averaging effects on PCD area.

*Microbubbles (Bracco BG9732)* – In this study, we used monodisperse MBs (BG9732, 6 μm diameter, Bracco). The geometric standard deviation, as provided by the manufacturer, was 1.15, and the experimentally determined polydispersity index was approximately 7%. Our monodisperse MBs had a mean MB concentration of 130 millions/mL, and volume of 15 μL/mL ($\beta = 1.5\%$ undiluted). The MB samples were initially freeze-dried; for resuspension, we followed the manufacturer's instruction where we added saline (0.9% Sodium Chloride). To take samples from the vial, we inserted two needles: one to sample the MB and one to balance the internal pressure. Sampled MBs were immediately used. In our in vitro setup, we diluted the MBs 100 times in degassed water, to match the void fraction ($\beta$) to 0.015%.

*High framerate microscopy* – To validate LAWPS algorithm, we incorporated custom designed optical high framerate microscopy setup. We attached laser cut acrylic plates to a box-like formation, and at the bottom we attached ibidi μ-slide I Luer (ibidi, Germany), 200 μm channel height, with the thinner wall facing downwards (thick wall facing US field). At the sides, we laser cut a hole to the acrylic plate to place A305 transducer (0.5 MHz center frequency, Olympus) and a 6 mm passive cavitation detector (3.5 MHz center frequency, Imasonic), both of which directly faced the center of ibidi channel. A305 transducer was positioned 65.5 mm from the ibidi channel, and the passive cavitation detector was positioned 30 mm from the ibidi channel. Before MB injection, ibidi slide was flushed and incubated with bovine serum albumin (BSA) under room temperature to prevent potential nonspecific binding of MBs onto the ibidi slide [53]. The acrylic tank was filled with degassed water prior to experiments.

Before injecting the MBs into the ibidi slide, we recorded a background AE using the same pressure (e.g., 100-250 kPa) that was used with MBs. Monodisperse MBs (BG9732, 6 μm diameter, Bracco) were then diluted 100 times (ideal $\beta = 0.015\%$, as well as distance between MBs > 15 $R_0$ as confirmed optically) and injected into ibidi slide. After injection, we waited for 1 minute to let MBs float to top of ibidi channel. We then sonicated the MB field with 30 cycle 0.5 MHz with varying pressures and recorded AE and high framerate video (5 Mfps, 256 frames – 51 μs video, Shimadzu) simultaneously. Just before sonication we recorded a large field of view image to be used in the analysis for number of MBs in the field (**Fig. S4**). Also to locate any MB translation due to primary radiation force or secondary Bjerknes forces, we imaged in high magnification before and after sonication (**Fig. S4**). MBs were slightly flushed after a recording session to replenish the MBs in the field.

*Optical analysis of MB dynamics* – After the acquisition of both the AE and the optical high framerate data, we analyzed the optical data (256 frame video) by first processing each frame with edge detection. Specifically, we first created separate videos of each MBs in the field of view (1 pixel approximates to 0.5 μm). We then applied edge detection to the separated videos to isolate MB boundary. To the MB boundary, we calculated diameter at each frame by measuring the furthest points of intersection, which were repeated for 359 degrees rotation. Mean diameter at a specific frame was calculated by taking the mean of 360 diameter measurements around the MB. The MB radius time series for each MB was calculated by analyzing all 256 frames for each MB.

*Acoustic analysis of MB dynamics* – The AE signals recorded from each session were first processed with background (fundamental reflections) removal. We subtracted the background signal without MBs from the AE signal with MBs, by aligning them in time domain using cross correlation. We then applied appropriate multipliers such as DI/CI ratio (for the source/receiver alignment and harmonic frequencies), number of MBs (manually counted from large field of view image acquired from optical microscopy), frequency-dependent sensitivity of PCD, and losses due to reflections/attenuation from ibidi channel (see next section) onto the isolated MB signal. Then, the resulting signal was assumed to be the signal that represents average MB dynamics in the US field. After converting the amplitude of signal into pressure using PCD calibration, LAWPS backpropagation algorithm was applied to estimate the average MB radius dynamics.

*k-wave simulation of ibidi channel* – We utilized the k-wave tool [54] in MATLAB to simulate the losses of AE propagation through the ibidi slide. The simulated field was a 50 mm (height) x 20 mm (width) 2D field, with grid size of 100 μm. Throughout the longer axis, we set the layers starting from air (10 mm), bottom of ibidi slide (180 μm), ibidi channel (1.65 mm), and top of ibidi slide (200 μm), and rest of field was water. Acoustic properties of each layer were assumed to be following; air – speed of sound 343 m/s, density 1.225 kg/m$^3$, ibidi slide top/bottom – speed of sound 2318 m/s [53], density 970 kg/m$^3$, ibidi channel – speed of sound 1500 m/s, density 998 kg/m$^3$, water – speed of sound 1500 m/s, density 998 kg/m$^3$. We then used poisson disk method to randomly locate 10 monopoles within the ibidi channel (20mm x 200 μm). We repeated the process for 100 different random locations of monopoles oscillating at 0.5 MHz and analyzed the recorded loss compared to the case where ibidi slide and air were absent. We observed that for 100 different simulations, the mean loss was approximately 6%, possibly due to combined effect of monopole's behavior next to hard/soft boundaries and reflection/transmission.

*Calibration of pressures* – We calibrated A305 transducer (0.5 MHz center frequency, 33 mm focal distance, Olympus) with a needle hydrophone (2mm model, Precision Acoustics) in free field water tank. The pressures reported in this study are peak to peak pressure in free field. To calibrate PCD that was used with high framerate microscopy (3.5 MHz center frequency, Imasonic), we precisely placed the PCD and the needle hydrophone at 65.5 mm from A305 transducer by calculating US travel distance. We chose 65.5 mm (approximately twice the focal distance) to mimic plane waves. At this location, the beam width (full width half maximum) was determined to be 10 mm. PCD's sensitivity at 0.5, 1, 1.5, 2, 2.5 MHz was determined to be 3.3, 4.5, 5.7, 5.6, and 8.7 mV/kPa, which is comparable to needle hydrophone's sensitivity (4.3, 3.5, 3.3, 3.1, 3.1, 3.0 mV/kPa at those frequencies). The calibration of PCD was used to convert the AE signals into pressure in the high framerate microscope setup.

## Acknowledgments

**General:** Hohyun Lee and Costas Arvanitis are inventors on a patent application (US Application No. 19/327,470) related to the work presented in this manuscript.

**Funding:** This study was supported by the NIH Grant R37CA239039 (NCI). This research was additionally supported by the Korea Health Technology R&D Project through the Korea Health Industry Development Institute, funded by the Ministry of Health & Welfare, Republic of Korea (Grant number: RS-2023-KH135060, and RS-2024-00337463)

**CRediT authorship contribution statement:**

**Hohyun Lee**: Writing – review & editing, Writing – original draft, Validation, Methodology, Conceptualization, Software. **Reza Pakdaman Zangabad**: Writing – review & editing, Methodology. **Chulyong Kim**: Writing – review & editing, Methodology. **Victor Menezes**: Writing – review & editing, Methodology. **Levent F. Degertekin**: Writing – review & editing,

Methodology. **Costas Arvanitis**: Writing – review & editing, Funding acquisition, Resources, Conceptualization, Project administration.

**Competing interests:** The authors declare no competing interests.

**Data and materials availability:** All data needed to evaluate the conclusions in the paper are present in the paper and/or the Supplementary Materials.

**References**

[1] S. Qin, C.F. Caskey, K.W. Ferrara, Ultrasound contrast microbubbles in imaging and therapy: physical principles and engineering, Phys. Med. Biol. 54 (2009) R27. https://doi.org/10.1088/0031-9155/54/6/R01.
[2] A.L. Klibanov, Microbubble Contrast Agents: Targeted Ultrasound Imaging and Ultrasound-Assisted Drug-Delivery Applications, Invest. Radiol. 41 (2006) 354. https://doi.org/10.1097/01.rli.0000199292.88189.0f.
[3] H. Estrada, T. Deffieux, J. Robin, M. Tanter, D. Razansky, Imaging the brain by traversing the skull with light and sound, Nat. Biomed. Eng. (2025) 1–17. https://doi.org/10.1038/s41551-025-01433-5.
[4] C.D. Arvanitis, N. Vykhodtseva, F. Jolesz, M. Livingstone, N. McDannold, Cavitation-enhanced nonthermal ablation in deep brain targets: feasibility in a large animal model, J. Neurosurg. 124 (2015) 1450–1459. https://doi.org/10.3171/2015.4.JNS142862.
[5] I. Lentacker, I. De Cock, R. Deckers, S.C. De Smedt, C.T.W. Moonen, Understanding ultrasound induced sonoporation: Definitions and underlying mechanisms, Adv. Drug Deliv. Rev. 72 (2014) 49–64. https://doi.org/10.1016/j.addr.2013.11.008.
[6] E. Stride, C. Coussios, Nucleation, mapping and control of cavitation for drug delivery, Nat. Rev. Phys. 1 (2019) 495–509. https://doi.org/10.1038/s42254-019-0074-y.
[7] A. Alva, C. Kim, P. Premdas, Y. Ferry, H. Lee, N. Lal, B. Jing, E. Botchwey, B.D. Lindsey, C. Arvanitis, Imaging of macrophage accumulation in solid tumors with ultrasound, Nat. Commun. 16 (2025) 6322. https://doi.org/10.1038/s41467-025-61624-1.
[8] Y. Guo, H. Lee, C. Kim, C. Park, A. Yamamichi, P. Chuntova, M.O. Bernabeu, H. Okada, H. Jo, C. Arvanitis, Ultrasound frequency-controlled microbubble dynamics in brain vessels regulate the enrichment of inflammatory pathways in the blood-brain barrier, Nat. Commun. in press (2024).
[9] N. de Jong, P.J.A. Frinking, A. Bouakaz, M. Goorden, T. Schourmans, X. Jingping, F. Mastik, Optical imaging of contrast agent microbubbles in an ultrasound field with a 100-MHz camera, Ultrasound Med. Biol. 26 (2000) 487–492. https://doi.org/10.1016/S0301-5629(99)00159-3.
[10] M. Postema, A. van Wamel, C.T. Lancée, N. de Jong, Ultrasound-induced encapsulated microbubble phenomena, Ultrasound Med. Biol. 30 (2004) 827–840. https://doi.org/10.1016/j.ultrasmedbio.2004.02.010.
[11] V. Garbin, D. Cojoc, E. Ferrari, E. Di Fabrizio, M.L.J. Overvelde, S.M. van der Meer, N. de Jong, D. Lohse, M. Versluis, Changes in microbubble dynamics near a boundary revealed by combined optical micromanipulation and high-speed imaging, Appl. Phys. Lett. 90 (2007) 114103. https://doi.org/10.1063/1.2713164.
[12] P. Marmottant, S. van der Meer, M. Emmer, M. Versluis, N. de Jong, S. Hilgenfeldt, D. Lohse, A model for large amplitude oscillations of coated bubbles accounting for buckling and rupture, J. Acoust. Soc. Am. 118 (2005) 3499–3505. https://doi.org/10.1121/1.2109427.
[13] M. Cattaneo, G. Guerriero, G. Shakya, L.A. Krattiger, L. G. Paganella, M.L. Narciso, O. Supponen, Cyclic jetting enables microbubble-mediated drug delivery, Nat. Phys. 21 (2025) 590–598. https://doi.org/10.1038/s41567-025-02785-0.
[14] A. Alva, C. Kim, P. Premdas, Y. Ferry, H. Lee, N. Lal, B. Jing, E. Botchwey, B.D. Lindsey, C. Arvanitis, Imaging of macrophage accumulation in solid tumors with ultrasound, Nat. Commun. 16 (2025) 6322. https://doi.org/10.1038/s41467-025-61624-1.
[15] T. Faez, I. Skachkov, M. Versluis, K. Kooiman, N. de Jong, In vivo Characterization of Ultrasound Contrast Agents: Microbubble Spectroscopy in a Chicken Embryo, Ultrasound Med. Biol. 38 (2012) 1608–1617. https://doi.org/10.1016/j.ultrasmedbio.2012.05.014.

[16] J.H. Bezer, P. Prentice, W.L.K. Chang, S.V. Morse, K. Christensen-Jeffries, C.J. Rowlands, A.S. Kozlov, J.J. Choi, Microbubble dynamics in brain microvessels, PLOS ONE 20 (2025) e0310425. https://doi.org/10.1371/journal.pone.0310425.

[17] S. Schoen, M.S. Kilinc, H. Lee, Y. Guo, F.L. Degertekin, G.F. Woodworth, C. Arvanitis, Towards controlled drug delivery in brain tumors with microbubble-enhanced focused ultrasound, Adv. Drug Deliv. Rev. 180 (2022) 114043. https://doi.org/10.1016/j.addr.2021.114043.

[18] H. Lee, Y. Guo, J.L. Ross, S. Schoen, F.L. Degertekin, C. Arvanitis, Spatially targeted brain cancer immunotherapy with closed-loop controlled focused ultrasound and immune checkpoint blockade, Sci. Adv. 8 (2022) eadd2288. https://doi.org/10.1126/sciadv.add2288.

[19] Y. Guo, H. Lee, Z. Fang, A. Velalopoulou, J. Kim, M.B. Thomas, J. Liu, R.G. Abramowitz, Y. Kim, A.F. Coskun, D.P. Krummel, S. Sengupta, T.J. MacDonald, C. Arvanitis, Single-cell analysis reveals effective siRNA delivery in brain tumors with microbubble-enhanced ultrasound and cationic nanoparticles, Sci. Adv. 7 (2021) eabf7390. https://doi.org/10.1126/sciadv.abf7390.

[20] C.D. Arvanitis, M.S. Livingstone, N. Vykhodtseva, N. McDannold, Controlled Ultrasound-Induced Blood-Brain Barrier Disruption Using Passive Acoustic Emissions Monitoring, PLOS ONE 7 (2012) 16.

[21] N. McDannold, N. Vykhodtseva, K. Hynynen, Targeted disruption of the blood–brain barrier with focused ultrasound: association with cavitation activity, Phys. Med. Biol. 51 (2006) 793–807. https://doi.org/10.1088/0031-9155/51/4/003.

[22] M.D. Gray, E. Lyka, C.C. Coussios, Diffraction Effects and Compensation in Passive Acoustic Mapping, IEEE Trans. Ultrason. Ferroelectr. Freq. Control 65 (2018) 258–268. https://doi.org/10.1109/TUFFC.2017.2778509.

[23] K.J. Haworth, K.B. Bader, K.T. Rich, C.K. Holland, T.D. Mast, Quantitative Frequency-Domain Passive Cavitation Imaging, IEEE Trans. Ultrason. Ferroelectr. Freq. Control 64 (2017) 177–191. https://doi.org/10.1109/TUFFC.2016.2620492.

[24] G. Renaud, J.G. Bosch, A.F.W. van der Steen, N. de Jong, An "acoustical camera" for in vitro characterization of contrast agent microbubble vibrations, Appl. Phys. Lett. 100 (2012) 101911. https://doi.org/10.1063/1.3693522.

[25] J. Sijl, H.J. Vos, T. Rozendal, N. de Jong, D. Lohse, M. Versluis, Combined optical and acoustical detection of single microbubble dynamics, J. Acoust. Soc. Am. 130 (2011) 3271–3281. https://doi.org/10.1121/1.3626155.

[26] M. Krack, J. Gross, Harmonic Balance for Nonlinear Vibration Problems, Springer International Publishing, Cham, 2019. https://doi.org/10.1007/978-3-030-14023-6.

[27] P. Deuflhard, Newton Methods for Nonlinear Problems: Affine Invariance and Adaptive Algorithms, Springer Berlin Heidelberg, Berlin, Heidelberg, 2011. https://doi.org/10.1007/978-3-642-23899-4.

[28] T.G. Leighton, 1 - The Sound Field, in: T.G. Leighton (Ed.), Acoust. Bubble, Academic Press, 1994: pp. 1–66. https://doi.org/10.1016/B978-0-12-441920-9.50006-7.

[29] L.E. Kinsler, A.R. Frey, A.B. Coppens, J.V. Sanders, Fundamentals of Acoustics, John Wiley & Sons, 2000.

[30] L. Hoff, P.C. Sontum, J.M. Hovem, Oscillations of polymeric microbubbles: Effect of the encapsulating shell, J. Acoust. Soc. Am. 107 (2000) 2272–2280. https://doi.org/10.1121/1.428557.

[31] K. Vokurka, On Rayleigh's model of a freely oscillating bubble. I. Basic relations, Czechoslov. J. Phys. 35 (1985) 28–40. https://doi.org/10.1007/BF01590273.

[32] T.G. Leighton, 4 - The Forced Bubble, in: T.G. Leighton (Ed.), Acoust. Bubble, Academic Press, 1994: pp. 287–438. https://doi.org/10.1016/B978-0-12-441920-9.50009-2.

[33] P. Marmottant, S. Hilgenfeldt, Controlled vesicle deformation and lysis by single oscillating bubbles, Nature 423 (2003) 153–156. https://doi.org/10.1038/nature01613.

[34] S.M. Nejad, H. Hosseini, H. Akiyama, K. Tachibana, Reparable Cell Sonoporation in Suspension: Theranostic Potential of Microbubble, Theranostics 6 (2016) 446–455. https://doi.org/10.7150/thno.13518.

[35] S.J. Lighthill, Acoustic streaming, J. Sound Vib. 61 (1978) 391–418. https://doi.org/10.1016/0022-460X(78)90388-7.

[36] P. Marmottant, T. Biben, S. Hilgenfeldt, Deformation and rupture of lipid vesicles in the strong shear flow generated by ultrasound-driven microbubbles, Proc. R. Soc. Math. Phys. Eng. Sci. 464 (2008) 1781–1800. https://doi.org/10.1098/rspa.2007.0362.

[37] A. Pommella, N.J. Brooks, J.M. Seddon, V. Garbin, Selective flow-induced vesicle rupture to sort by membrane mechanical properties, Sci. Rep. 5 (2015) 13163. https://doi.org/10.1038/srep13163.
[38] R. Koda, T. Origasa, T. Nakajima, Y. Yamakoshi, Observing Bubble Cavitation by Back-Propagation of Acoustic Emission Signals, IEEE Trans. Ultrason. Ferroelectr. Freq. Control 66 (2019) 823–833. https://doi.org/10.1109/TUFFC.2019.2897983.
[39] T. Sun, Y. Zhang, C. Power, P.M. Alexander, J.T. Sutton, M. Aryal, N. Vykhodtseva, E.L. Miller, N.J. McDannold, Closed-loop control of targeted ultrasound drug delivery across the blood–brain/tumor barriers in a rat glioma model, Proc. Natl. Acad. Sci. 114 (2017) E10281–E10290. https://doi.org/10.1073/pnas.1713328114.
[40] R. Pakdaman Zangabad, H. Lee, X. Zhang, M. Sait Kilinc, C.D. Arvanitis, F. Levent Degertekin, A High Sensitivity CMUT-Based Passive Cavitation Detector for Monitoring Microbubble Dynamics During Focused Ultrasound Interventions, IEEE Trans. Ultrason. Ferroelectr. Freq. Control 71 (2024) 1087–1096. https://doi.org/10.1109/TUFFC.2024.3436918.
[41] M.S. Kilinc, H. Lee, C.D. Arvanitis, F.L. Degertekin, Dual mode CMUT Array Operation for Skull Imaging and Passive Acoustic Monitoring in Transcranial Ultrasound, in: 2021 IEEE Int. Ultrason. Symp. IUS, 2021: pp. 1–4. https://doi.org/10.1109/IUS52206.2021.9593365.
[42] M.S. Kilinc, H. Lee, Y.R. Ferry, B. Ingram, B. Skowronski, P.P. Dash, R.P. Zangabad, C. Arvanitis, F.L. Degertekin, A Piezo-Cmut Hybrid Hemispherical Transmit Array for Passive Acoustic Mapping of Microbubble Activity, in: 2025 IEEE Int. Ultrason. Symp. IUS, 2025: pp. 1–5. https://doi.org/10.1109/IUS62464.2025.11201469.
[43] J. Wu, Acoustic Streaming and Its Applications, Fluids 3 (2018). https://doi.org/10.3390/fluids3040108.
[44] O. Couture, V. Hingot, B. Heiles, P. Muleki-Seya, M. Tanter, Ultrasound Localization Microscopy and Super-Resolution: A State of the Art, IEEE Trans. Ultrason. Ferroelectr. Freq. Control 65 (2018) 1304–1320. https://doi.org/10.1109/TUFFC.2018.2850811.
[45] L.A. Crum, Bjerknes forces on bubbles in a stationary sound field, J. Acoust. Soc. Am. 57 (1975) 1363–1370. https://doi.org/10.1121/1.380614.
[46] Z. Kou, Q. You, J. Kim, Z. Dong, M.R. Lowerison, N.V.C. Sekaran, D.A. Llano, P. Song, M.L. Oelze, High-Level Synthesis Design of Scalable Ultrafast Ultrasound Beamformer With Single FPGA, IEEE Trans. Biomed. Circuits Syst. 17 (2023) 446–457. https://doi.org/10.1109/TBCAS.2023.3267614.
[47] C. Yu, Y. Su, J. Lee, K. Chai, B. Kim, A 32x32 Time-Domain Wavefront Computing Accelerator for Path Planning and Scientific Simulations, in: 2021 IEEE Cust. Integr. Circuits Conf. CICC, 2021: pp. 1–2. https://doi.org/10.1109/CICC51472.2021.9431448.
[48] J. Lee, K.Y. Huh, D. Kang, J. Lim, B.C. Lee, B. Lee, A low-complexity and high-frequency ASIC transceiver for an ultrasound imaging system, Biomed. Eng. Lett. 14 (2024) 1377–1384. https://doi.org/10.1007/s13534-024-00411-1.
[49] H. Lee, C. Arvanitis, Acoustical method for microbubble oscillatory radius estimation and acoustic emission detection, US20250076258A1, 2025. https://patents.google.com/patent/US20250076258A1/en (accessed September 19, 2025).
[50] R. Bridson, Fast Poisson disk sampling in arbitrary dimensions, in: ACM SIGGRAPH 2007 Sketches, ACM, San Diego California, 2007: p. 22. https://doi.org/10.1145/1278780.1278807.
[51] K.W. Commander, A. Prosperetti, Linear pressure waves in bubbly liquids: Comparison between theory and experiments, J. Acoust. Soc. Am. 85 (1989) 732–746. https://doi.org/10.1121/1.397599.
[52] P.J. Clark, F.C. Evans, Distance to Nearest Neighbor as a Measure of Spatial Relationships in Populations, Ecology 35 (1954) 445–453. https://doi.org/10.2307/1931034.
[53] R. Pakdaman Zangabad, H. Li, J.J.P. Kouijzer, S.A.G. Langeveld, I. Beekers, M. Verweij, N. De Jong, K. Kooiman, Ultrasonic Characterization of Ibidi μ-Slide I Luer Channel Slides for Studies With Ultrasound Contrast Agents, IEEE Trans. Ultrason. Ferroelectr. Freq. Control 70 (2023) 422–429. https://doi.org/10.1109/TUFFC.2023.3250202.
[54] B.E. Treeby, B.T. Cox, k-Wave: MATLAB toolbox for the simulation and reconstruction of photoacoustic wave fields, J. Biomed. Opt. 15 (2010) 021314. https://doi.org/10.1117/1.3360308.

# Supporting Information

**Quantitative acoustic monitoring of ensembles of weakly nonlinear microbubble oscillations in optically inaccessible environments**

Hohyun Lee,[1] Reza Pakdaman Zangabad,[1] Chulyong Kim,[1] Victor Menezes[1], Juyoung Park[3,4], F. Levent Degertekin,[1] and Costas Arvanitis[1,2]*

**Affiliations**
[1] Georgia Institute of Technology, G.W. Woodruff School of Mechanical Engineering, Atlanta, USA
[2] Georgia Institute of Technology and Emory University, Wallace H. Coulter Department of Biomedical Engineering, Atlanta, USA
[3] Department of Biomedical Device, Gachon University College of IT Convergence, 1342 Seongnam-daero, Sujeong-gu, Seongnam 13120, Republic of Korea
[4] Neumous Inc., Osongsaengmyeong 1-ro, Osong-eup, Heungdeok-gu, Cheongju 28160, Republic of Korea

*Corresponding author
Costas Arvanitis (costas.arvanitis@gatech.edu)

**Experimental Section**

1. **Linear Acoustic Wave Propagation and Superposition (LAWPS) method using sub-harmonic multiples and fundamental multiples onto highly nonlinear microbubble oscillations**

We generated acoustic emissions (AEs) from highly nonlinear microbubble (MB) oscillations using Vokurka's method [1] and Linear Acoustic Wave Propagation and Superposition (LAWPS) method in propagation mode. The excitation pressure to MB model [2] was applied up to 200 kPa at 0.5 MHz frequency. For LAWPS algorithm, we compared effect of sub- and ultra- harmonics by using LAWPS operating at sub-harmonic multiple and fundamental multiples. At higher oscillations ($\Delta R/R_0 = 20\% \sim 40\%$), transient inertial responses from MBs were present, which were not effectively captured with Fourier series even with sub- and ultra-harmonics considered. While we found that adding sub- and ultra-harmonic components reduced errors in both LAWPS propagation and backpropagation, its effect was less than 1% improvement at lower oscillations, possibly due to lack of sub-harmonic oscillations.

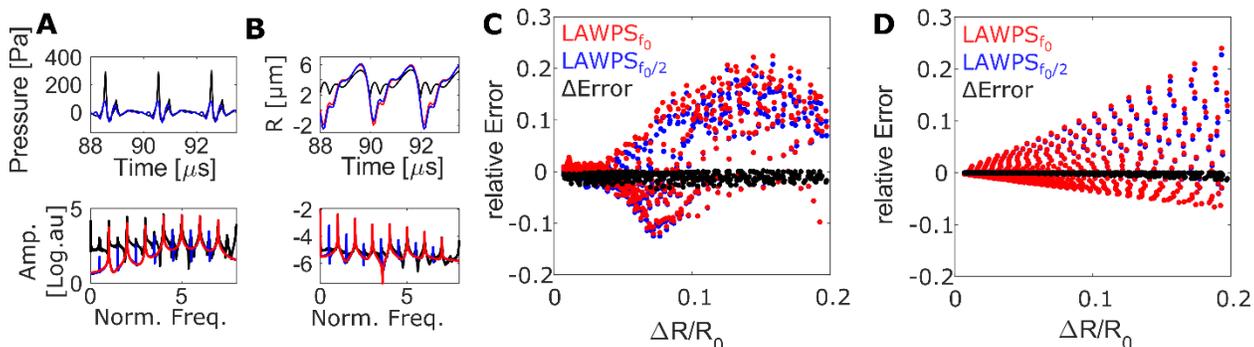

**Figure S1.** Sub-harmonic based LAWPS. A) AE propagation using Vokurka's method (black), fundamental (0.5 MHz)-based LAWPS (red), and sub-harmonic based LAWPS (blue). B) MB oscillation reconstruction using fundamental (0.5 MHz)-based LAWPS (red), and sub-harmonic

based LAWPS (blue). C) Relative error in AE propagation as a function of radius change. D) Relative error in MB radius backpropagation as a function of radius change.

## 2. Calibration of A305 focused ultrasound (FUS) transducer and passive cavitation detector (PCD)

We calibrated an A305 transducer (0.5 MHz center frequency, 33 mm focal distance, Olympus) with a needle hydrophone (2mm model, Precision Acoustics) in free field water tank. The pressures reported in this study are peak negative pressure in free field. To calibrate PCD that was used with high framerate microscopy (3.5 MHz center frequency, Imasonic), we precisely placed the PCD and the needle hydrophone at 65.5 mm from A305 transducer by calculating US travel distance. We chose 65.5 mm (approximately twice the focal distance) to mimic plane waves. The calibration was used to convert the AE signals into pressure in the high framerate microscope setup.

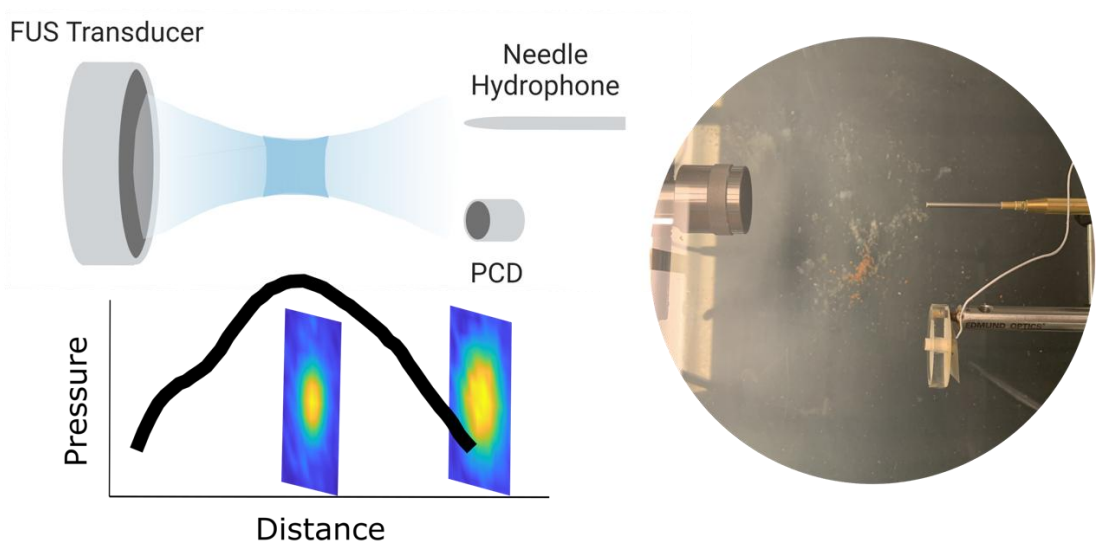

**Figure S2.** Calibration of FUS transducer and PCD.

## 3. k-wave simulation of ibidi microfluidic device

We utilized the k-wave tool [3] in MATLAB to simulate the losses of acoustic emission (AE) propagation through the ibidi slide. The simulated field was a 50 mm (height) x 20 mm (width) 2D field, with grid size of 100 μm. Throughout the longer axis, we set the layers starting from air (10 mm), bottom of ibidi slide (180 μm), ibidi channel (200 μm), and top of ibidi slide (1.65 mm), and rest of field was water. Acoustic properties of each layer were assumed to be following; air – speed of sound 343 m/s, density 1.225 kg/m$^3$, ibidi slide top/bottom – speed of sound 2318 m/s [4], density 970 kg/m$^3$, ibidi channel – speed of sound 1500 m/s, density 998 kg/m$^3$, water – speed of sound 1500 m/s, density 998 kg/m$^3$. We then used poisson disk method to randomly locate 10 monopoles within the ibidi channel (20mm x 200 μm). We repeated the process for 100 different random locations of monopoles oscillating at 0.5 MHz and analyzed the recorded loss compared to the case where ibidi slide and air were absent. We observed that for 100 different simulations, the mean loss was approximately 6%, possibly due to combined effect of monopole's behavior next to hard/soft boundaries and reflection/transmission.

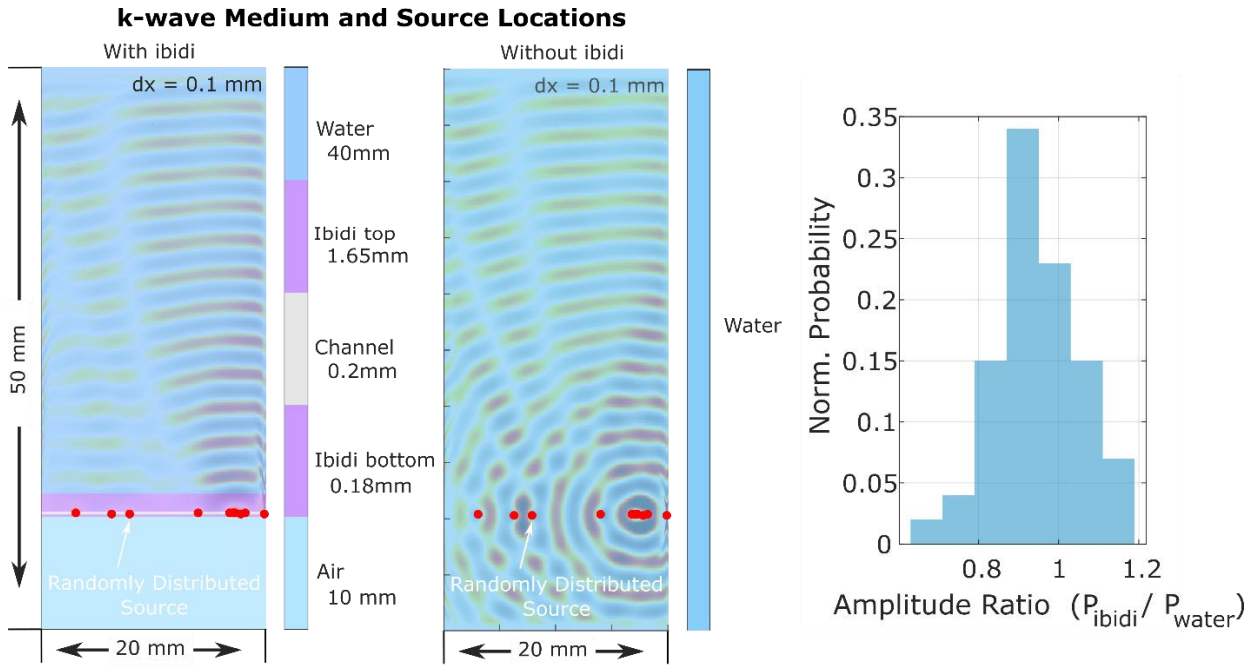

**Figure S3.** k-wave simulation of monopole sources in ibidi microslide. Left – with ibidi, right – without ibidi. Mean loss was 5.5%.

4. Microscopy image acquisition and analysis

To verify MB spatial distribution and motion before and after sonication, we performed bright-field microscopy using 20x and 60x objective lenses. The 20x images provided a large field of view for quantifying MB concentration and ensuring sufficient inter-MB spacing. High-magnification images of the same field were subsequently captured immediately before and after US exposure to track potential MB translation. Images were acquired at consistent focal planes. While small centroid shifts were detected after sonication, their magnitude is negligible compared with wavelength (3 mm). Thus, MBs remained as point sources and linear/monopole-based assumptions of LAWPS still hold.

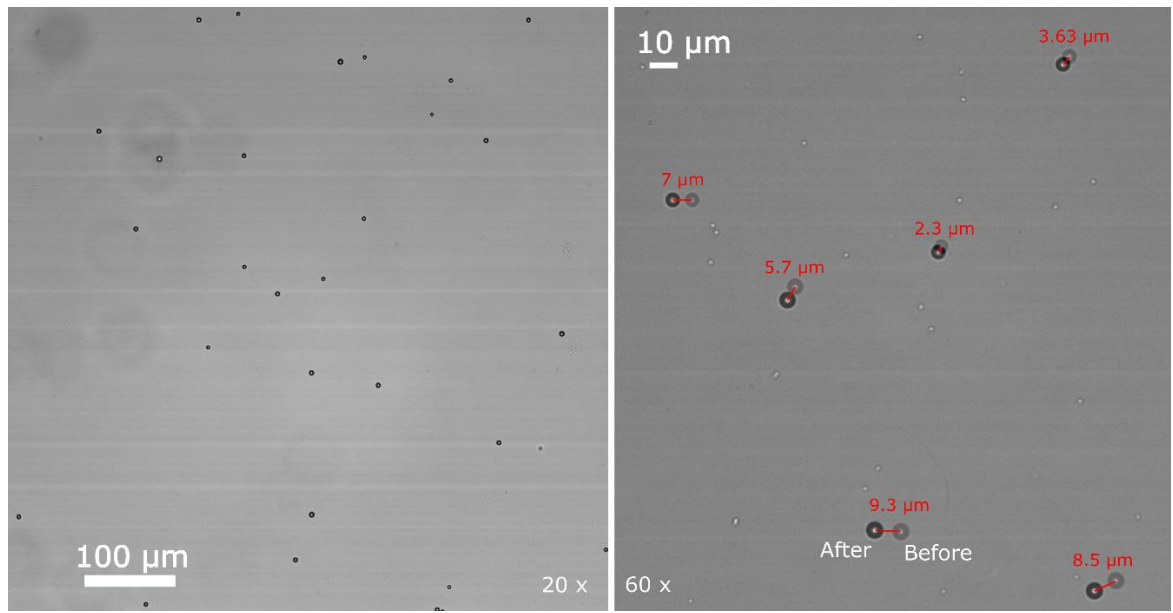

**Figure S4.** Optical validation of MB spatial distribution and pre- and post- sonication images. Left: large field of view microscope image used to confirm inter-MB spacing. The same image was used to estimate number of MBs in the focus. Right: high-magnification images of the same field of view acquired before and after US exposure. Red markers denote MB centroid displacements (23 ~ 9.3 µm) corresponding to translation induced by acoustic radiation and secondary Bjerknes forces during sonication.

## 5. Sub-harmonic components observed in Acoustic measurements

During our *in vitro* experiments, we observed sub-harmonic components in the acoustic emissions. However, these sub-harmonic components were not evident in optical measurements, which suggests that there may be weak asymmetric behavior that was not captured due to our optical resolution.

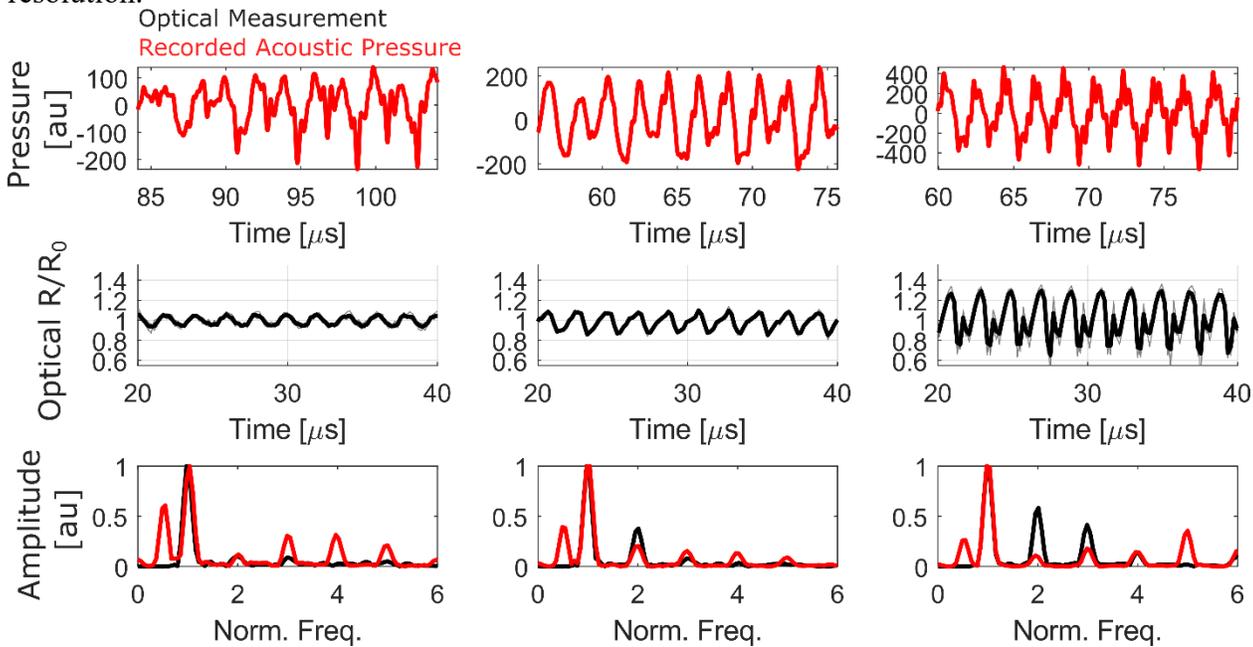

**Figure S5.** Frequency spectrum of acoustic and optical measurements. Top: Acoustic measurements. Middle: Optical measurements. Bottom: frequency spectrum of optical (black) and acoustic (red) measurements.

**References**
[1] K. Vokurka, On Rayleigh's model of a freely oscillating bubble. I. Basic relations, Czech J Phys 35 (1985) 28–40. https://doi.org/10.1007/BF01590273.
[2] L. Hoff, P.C. Sontum, J.M. Hovem, Oscillations of polymeric microbubbles: Effect of the encapsulating shell, The Journal of the Acoustical Society of America 107 (2000) 2272–2280. https://doi.org/10.1121/1.428557.
[3] B.E. Treeby, B.T. Cox, k-Wave: MATLAB toolbox for the simulation and reconstruction of photoacoustic wave fields, JBO 15 (2010) 021314. https://doi.org/10.1117/1.3360308.
[4] R. Pakdaman Zangabad, H. Li, J.J.P. Kouijzer, S.A.G. Langeveld, I. Beekers, M. Verweij, N. De Jong, K. Kooiman, Ultrasonic Characterization of Ibidi µ-Slide I Luer Channel Slides for Studies With Ultrasound Contrast Agents, IEEE Transactions on Ultrasonics, Ferroelectrics, and Frequency Control 70 (2023) 422–429. https://doi.org/10.1109/TUFFC.2023.3250202.